\documentclass{article} 
\usepackage[preprint]{colm2026_conference}

\usepackage{microtype}
\usepackage{hyperref}
\usepackage{url}
\usepackage{booktabs}

\usepackage{lineno}

\definecolor{darkblue}{rgb}{0, 0, 0.5}
\hypersetup{colorlinks=true, citecolor=darkblue, linkcolor=darkblue, urlcolor=darkblue}

\usepackage{times}
\usepackage{latexsym}

\usepackage[T1]{fontenc}

\usepackage[utf8]{inputenc}

\usepackage{microtype}

\usepackage{wrapfig} 

\usepackage{inconsolata}

\usepackage{graphicx}

\usepackage{enumitem}

\usepackage{subfig}
\usepackage{tablefootnote}
\usepackage{threeparttable}

\usepackage{algorithm}
\usepackage{algpseudocode}

\usepackage{graphicx}
\usepackage{tabularx}
\usepackage{threeparttable}
\usepackage{makecell}
\usepackage{multicol}
\usepackage{multirow}
\usepackage{xparse}
\usepackage{colortbl}

\usepackage{amsmath}
\usepackage{mathtools}
\usepackage{amsthm}

\usepackage{amsmath}
\usepackage{amssymb}
\usepackage{mathtools}
\usepackage{amsthm}
\usepackage{xparse}
\usepackage{xspace}
\usepackage{pifont}

\usepackage{listings}
\usepackage{balance}
\usepackage[most]{tcolorbox}

\lstdefinelanguage{json}{
    basicstyle=\ttfamily\small,  
    showstringspaces=false,      
    breaklines=true,             
    frame=none,                
    columns=fullflexible         
}

\usepackage{siunitx}
\usepackage{subcaption}

\usepackage[capitalize,noabbrev]{cleveref}

\theoremstyle{plain}
\newtheorem{theorem}{Theorem}[section]

\theoremstyle{definition}

\theoremstyle{remark}

\definecolor{myorange}{RGB}{199,93,24}

\title{\includegraphics[width=0.9cm, height=0.8cm]{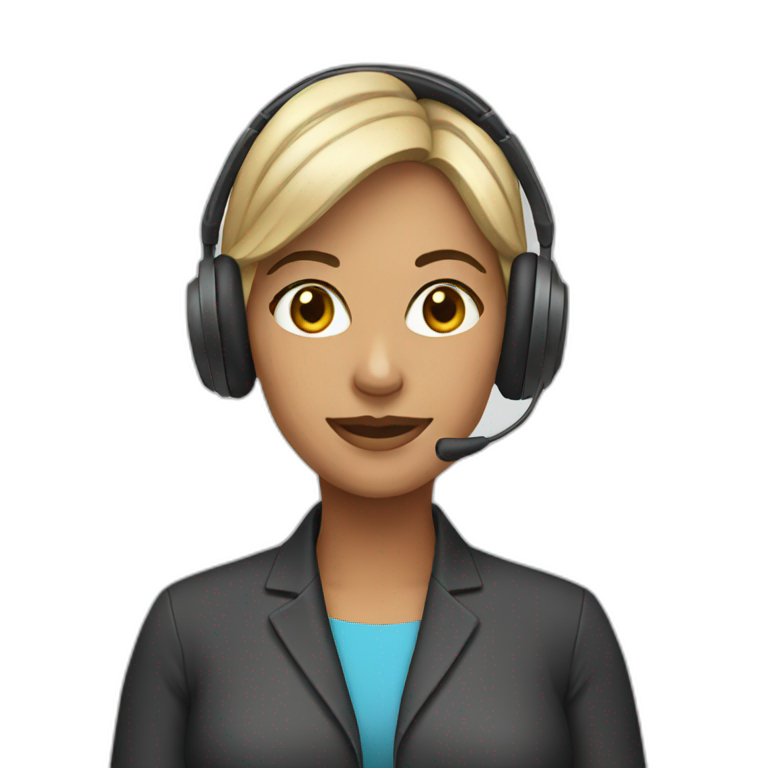} Talk to Right Specialists: Iterative Routing in Multi-agent Systems for Question Answering}

\author{Feijie Wu$^{\dag}$ \quad Zitao Li$^{\Delta}$\thanks{Work was done when ZL was employed at Alibaba} \quad Fei Wei$^{\ddag}$ \quad  Yaliang Li$^{\ddag}$ \quad Bolin Ding$^{\ddag}$ \quad Jing Gao$^{\dag}$ \\ 
    $^{\dag}$Purdue University \quad $^{\ddag}$Alibaba \quad $^{\Delta}$Zoom Communication \\
    \texttt{\{wu1977, jinggao\}@purdue.edu} \\
    \texttt{\{feiwei, yaliang.li, bolin.ding\}@alibaba-inc.com} \\
    \texttt{zitao.li@zoom.us}
}

\newcommand{\algo}{\ensuremath{\text{RIRS}}\xspace}

\renewcommand{\cite}{\citep}

\begin{document}

\ifcolmsubmission
\linenumbers
\fi

\maketitle
\begin{abstract}
Retrieval-augmented generation (RAG) agents are increasingly deployed to answer questions over local knowledge bases that cannot be centralized due to knowledge-sovereignty constraints. 
This results in two recurring failures in production: users do not know which agent to consult, and complex questions require evidence distributed across multiple agents. To overcome these challenges, we propose \algo, a training-free orchestration framework to enable a multi-agent system for question answering. In detail, \algo summarizes each agent's local corpus in an embedding space, enabling a user-facing server to route queries only to the most relevant agents, reducing latency and avoiding noisy "broadcast-to-all" contexts. For complicated questions, the server can iteratively aggregate responses to derive intermediate results and refine the question to bridge the gap toward a comprehensive answer. Extensive experiments demonstrate the effectiveness of \algo, including its ability to precisely select agents and provide accurate responses to single-hop queries, and its use of an iterative strategy to achieve accurate, multi-step resolutions for complex queries.
\end{abstract}

\section{Introduction}

\begin{figure*}
    \centering
    \includegraphics[width=\linewidth]{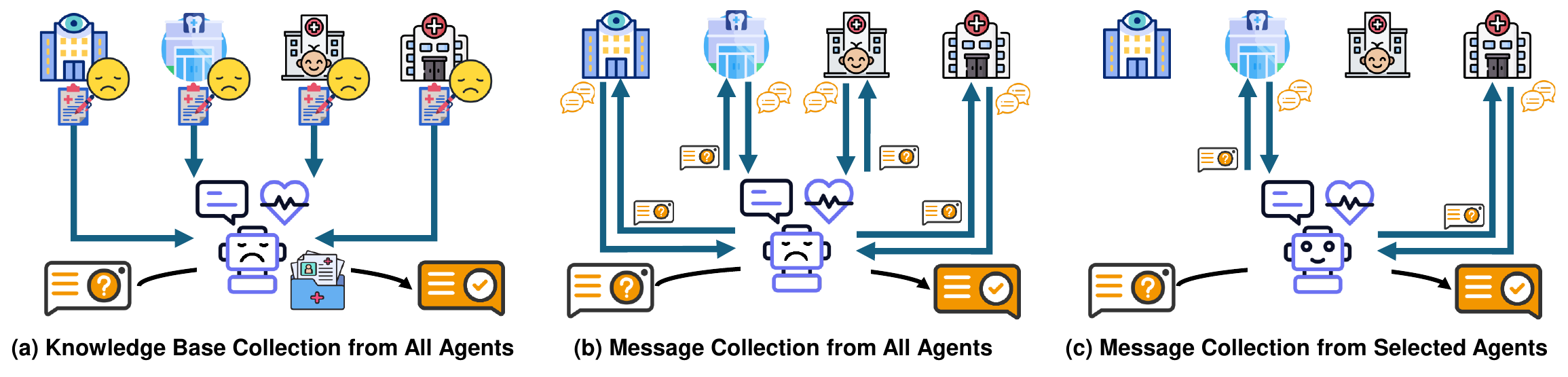}
    \subfloat{\label{subfig:(problem_a)}}{}
    \subfloat{\label{subfig:(problem_b)}}{}
    \subfloat{\label{subfig:(problem_c)}}{}
    \vspace{-20px}
    \caption{Collaboration Strategies of Multi-agent System for QA.}
    \vspace{-15px}
    \label{fig:multiagent_qa}
\end{figure*}

Retrieval-augmented generation (RAG) improves the reliability of large language model (LLM) applications by grounding model outputs in retrieved external evidence \cite{wu2024retrieval, asai2023retrieval, lewis2020retrieval, jiang2023active, izacard2020leveraging, mallen2022not, kasai2024realtime, xiong2024benchmarking, agent_intro, huggingfaceRAGAgent, joshi2024reaper}. In practice, this capability is often deployed as a \emph{RAG agent} backed by a private corpus. However, enterprise knowledge is rarely centralized: different business units, products, or teams often maintain separate RAG agents over disjoint corpora, such as internal documentation, contracts, incident reports, or partner data \cite{addison2024c, wu2024fedbiot, wu2024towards}. Centralizing these corpora into a single repository (\cref{subfig:(problem_a)}) is typically infeasible due to governance constraints, access controls, and confidentiality requirements. Moreover, as we show in \cref{subsec:single}, centralization is not only desirable due to knowledge sovereignty constraints, but also hurts retrieval quality and downstream reasoning. This setting raises a fundamental systems question: how can a user-facing server answer questions when the relevant evidence is distributed across multiple specialized RAG agents and cannot be pooled into a single repository? 

A natural orchestration strategy is \emph{broadcast-to-all}: as illustrated in \cref{subfig:(problem_b)}, a user-facing server forwards each query to every RAG agent, aggregates their responses, and repeats this process until a final answer is produced. In this system, each agent retains its local, domain-specific knowledge base, ensuring that its knowledge sovereignty is well protected and that its data are not merged into a centralized repository. However, broadcast-to-all is inefficient in practice: (i) it wastes computation and communication because many agents are irrelevant to a given query, (ii) it increases end-to-end latency by waiting for slow agents and processing long concatenated responses, and (iii) it can degrade answer quality because excessively long and noisy context makes it easy for the model to overlook the right evidence (the "lost-in-the-middle" problem), as demonstrated in \cref{subsec:all}. These limitations motivate a distributed multi-agent system with a \textit{routing mechanism} that selects only the most appropriate agents (treated as specialists) for each query, as illustrated in \cref{subfig:(problem_c)}.

Existing routing methods generally fall into two categories: description-based and learning-based. Description-based methods rely on manually written descriptions of agent capabilities, allowing an LLM to select the most suitable agent for a given query based on these descriptions \cite{patil2024gorilla, lu2024chameleon, lu2024gear, yuan2025easytool, wu2025portool, zheng2026skillrouter}. In contrast, learning-based methods train a router on task-specific data to direct each user query to the most appropriate agent \cite{lu2023routing, chen2024routerdc, ong2024routellm, yue2025masrouter, mei2025omnirouter, wei2025learning, wang2026icl}. However, both strategies are difficult to apply in multi-RAG-agent settings. A textual description may not accurately capture the knowledge boundary of an agent whose corpus covers multiple subtopics. Learning-based methods are also problematic because the effective capability of a RAG agent is determined by its non-parametric local corpus, not only by parametric knowledge, and it is difficult to construct an effective dataset that fully covers that corpus. Therefore, a routing mechanism in a multi-RAG-agent system should accurately characterize an agent's knowledge capabilities without requiring raw data sharing or expensive training.

In this work, we introduce \textbf{RIRS}, a practical orchestration framework that turns a collection of RAG agents into a single, coordinated problem-solving pipeline via \textit{\textbf{R}outing} and \textit{\textbf{I}terative \textbf{R}efining-\textbf{S}olving}. Specifically, this framework consists of two modules, namely, knowledge summarization and query solving. In the knowledge summarization stage, each agent encodes its local corpus, partitions the resulting embeddings into coherent clusters, and uploads only cluster centroids to the server. These centroids provide a compact embedding-space summary of the agent's knowledge capabilities, while keeping raw knowledge local. In the query-solving stage, the server routes each query to the most relevant agents by matching the query embedding to the uploaded centroids. For complex queries, the server then iteratively evaluates partial responses, refines the unresolved question, and continues querying only the agents needed for the remaining subproblem. In this way, RIRS avoids broadcast-to-all, preserves knowledge sovereignty, and supports evidence composition across agents and across rounds.

\vspace{-5pt}
\paragraph{Contributions.} We summarize the contributions of this paper as follows:
\begin{itemize}[topsep=0pt,itemsep=0pt,parsep=0pt,partopsep=0pt,leftmargin=*]
\item We study question answering in multi-RAG-agent systems where relevant evidence is distributed across specialized agents and cannot be centralized because of knowledge-sovereignty constraints. This setting exposes a key practical challenge: users do not know which agent to query, while complex questions often require evidence composition across agents. 
\item We propose \algo, a training-free framework built on two components: knowledge summarization, which represents each agent's local corpus through cluster-level embedding summaries, and query solving, which selectively routes queries, aggregates partial progress, and refines unresolved questions across rounds. We also provide a theoretical analysis showing that the routing error of the centroid-based summary is bounded under a cluster-coherence condition.
\item We show that \algo improves both routing quality and question answering performance. In particular, it enables more precise agent selection for single-hop queries and more accurate multi-step resolution for complex queries, outperforming representative baselines in both effectiveness and efficiency.
\end{itemize}

\vspace{-5pt}
\section{Related Work}\label{apdx:related_works}
\vspace{-5pt}

\paragraph{Retrieval Augmented Generation (RAG).} RAG has gained substantial interest in academic research as a robust framework that integrates external knowledge sources into large language models to enhance the quality and reliability of generated responses \cite{lewis2020retrieval, chen2017reading, guu2020retrieval, karpukhin2020dense, izacard2020leveraging, borgeaud2022improving, yu2022retrieval, shi2023replug, yan2024corrective, li2023llatrieval, press2022measuring, chan2024rq, su2024dragin, trivedi2022interleaving, ma2023query, shao2023enhancing, zhang2025belle}, and there is a detailed description of how an RAG agent works in Section \ref{subsec:prelim}. Recent work has increasingly moved toward agentic RAG, where the model actively plans retrieval, refines queries, critiques intermediate results, and iterates between retrieval and reasoning to handle complex questions more effectively \cite{asai2023self, jiang2023active, zhuang2024efficientrag, verma2024plan, yang2024rag, singh2025agentic, li2025towards}. These advances improve a single RAG pipeline, typically over a centralized knowledge source. In contrast, our work studies a distributed setting in which knowledge is partitioned across multiple RAG agents, so the main challenge is not only iterative reasoning, but also routing the query to the right agents and composing evidence across agents and rounds.

\vspace{-5pt}
\paragraph{Multi-agent Systems.} 
Multi-agent systems leverage multiple specialized LLM-based agents to collaboratively solve complex tasks, and have attracted growing attention in recent years \cite{hong2023metagpt, li2023camel, wu2023autogen, chen2023autoagents, chen2023agentverse, zhao2023competeai, guo2024large, chen2024mindsearch, li2024agent, ye2025mas, liu2026vision}. In question answering, a growing number of studies improve answer quality through multi-agent collaboration, such as debate- or discussion-based frameworks in which multiple agents reason over a shared or common knowledge source to refine candidate answers \cite{zhang2025agentrouter, iannelli2024sla, du2024improving, zhang2025belle}. These methods primarily emphasize better coordination and reasoning among agents once the knowledge source becomes accessible. Different from these works, our focus is not on having multiple agents reason over the same repository, but on identifying the most relevant agents when knowledge is inherently distributed across agents.

\vspace{-5pt}
\section{Preliminary and Motivations} \label{sec:prelim_motivation}
\vspace{-5pt}

This section first reviews the pipelines of a RAG agent and a multi-RAG-agent system (\cref{subsec:prelim}). We then motivate multi-agent design by showing that multi-RAG-agent systems can consistently outperform single-agent baselines (\cref{subsec:single}), and by highlighting why an effective routing mechanism is crucial (\cref{subsec:all}).

\vspace{-5pt}
\subsection{Preliminary} \label{subsec:prelim}

\vspace{-5pt}
\paragraph{RAG Agent.} 
An RAG agent is an advanced application that integrates retrieval mechanisms with LLMs to deliver accurate and contextually rich responses. Specifically, an RAG agent processes an input query through three steps: (i) \textit{Knowledge Retrieval:} It extracts relevant knowledge pieces using both sparse and dense retrievers from external sources \cite{robertson2004simple, izacard2021unsupervised, xu2023retrieval}; (ii) \textit{Reranking:} It ranks those retrieved pieces with a pretrained regression model to identify the likelihood of a knowledge piece helpful to the input query, thereby filtering out unhelpful or misleading information and allowing the generation model to focus on the most pertinent content; (iii) \textit{Response Generation:} It combines the question and the pertinent external knowledge to produce an informed response with its backbone LLM. This process is also known as RankRAG \cite{yu2024rankrag}. For multi-hop questions, it can adopt EfficientRAG \cite{zhuang2024efficientrag} to iteratively refine the query and repeat retrieval and generation until the question is resolved. 

\vspace{-5px}
\paragraph{Multi-RAG-Agent System.} 
A multi-RAG-agent system consists of numerous RAG agents orchestrated under a central entity (i.e., server). When a user query comes, the system should repeatedly follow these three steps until the query is resolved: (i) the server selects a group of agents and forwards the query to them; (ii) the chosen RAG agent should give a response to the best of its local knowledge; (iii) the server aggregates the responses from the selected agents and simplifies the queries. While answering a given query, an RAG agent is supposed not to expose local raw knowledge, thereby preserving agent privacy and knowledge sovereignty. Such a system is also known as the federated RAG system in some recent works \cite{addison2024c, shi2025privacy, jung2025federated, zhao2024frag}. 

\vspace{-5px}
\subsection{Single-RAG-Agent System Does NOT Work} \label{subsec:single}
\vspace{-5pt}

\begin{table}[t]
    \centering
    \caption{Performance comparison on Natural Questions (NQ) and HotpotQA. Best results under Qwen-Plus and LLaMA-3.1-8B are in \textbf{bold} and \underline{underline}, respectively, which will be applied to the rest of the paper without annotations.}
    \vspace{-10px}
    \definecolor{lightgray}{gray}{0.9}
    \renewcommand{\arraystretch}{1.2}
    \resizebox{\linewidth}{!}{
    \begin{threeparttable}
\begin{tabular}{l l | c c | c c c | c c c | c c c}
    \Xhline{1.5pt}
    \multirow{2}{*}{Methods} & \multirow{2}{*}{Models}
        & \multicolumn{2}{c|}{Natural Questions}
        & \multicolumn{3}{c|}{HotpotQA}
        & \multicolumn{3}{c|}{2WikiMultiHopQA}
        & \multicolumn{3}{c}{MuSiQue} \\\cline{3-13}
    & & LM & Acc.
        & LM & F1 & Acc.
        & LM & F1 & Acc.
        & LM & F1 & Acc.  \\
    \hline
        \multirow{2}{*}{\makecell[l]{CoT\\(Without RAG)}} 
            & Qwen-Plus    
                & 52.99 & 74.55
                & 39.68 & 43.82 & 54.58 
                & 42.38 & 44.89 & 37.31 
                & 16.76 & 20.66 & 33.86\\
            & LLaMA-3.1-8B    
                & 44.41 & 58.60
                & 30.96 & 36.02 & 44.23  
                & 31.40 & 34.46 & 25.99  
                & 13.30 & 17.76 & 21.86 \\
        \hline
    \multirow{2}{*}{\makecell[l]{EfficientRAG\\(Single-RAG-Agent)}} 
        & Qwen-Plus    
            & 62.73 & 82.61
            & 57.41 & 63.01 & 76.55
            & 69.41 & 71.77 & 75.85
            & 26.62 & 31.41 & 46.42 \\
        & LLaMA-3.1-8B    
            & \underline{56.48} & \underline{70.10}
            & 47.38 & 52.96 & 61.17
            & 48.45 & 52.04 & 51.13
            & 21.38 & 26.22 & 29.13 \\
    \hline
    \multirow{2}{*}{\makecell[l]{GoldRouter\\(Multi-RAG-Agent)}} 
        & Qwen-Plus    
            & \textbf{63.41} & \textbf{83.02}
            & \textbf{58.17} & \textbf{63.99} & \textbf{81.01}
            & \textbf{70.11} & \textbf{72.37} & \textbf{76.33}
            & \textbf{29.01} & \textbf{33.75} & \textbf{47.55} \\
        & LLaMA-3.1-8B    
            & 55.13 & 69.70
            & \underline{50.60} & \underline{56.26} & \underline{64.94}
            & \underline{49.73} & \underline{52.99} & \underline{52.07}
            & \underline{23.27} & \underline{28.18} & \underline{31.46} \\\Xhline{1.5pt}
\end{tabular}
    \end{threeparttable} 
    }
    \vspace{-15pt}
    \label{tab:single_multiple_cmp}
\end{table}

In a single-RAG-agent system (Figure \ref{subfig:(problem_a)}), one agent collects and stores all documents in a single repository. Such centralization is often impractical in deployments due to knowledge-sovereignty concerns. Additionally, we analyze its performance limitation by comparing it with a practical multi-RAG-agent system. Table \ref{tab:single_multiple_cmp} compares three baselines that isolate retrieval and agent organization. \textit{Chain-of-Thought (CoT)} uses only the LLM's parametric knowledge. \textit{GoldRouter} is an oracle upper bound for a multi-RAG-agent system that always routes each step to the relevant agents and aggregates their evidence.

The results in the table confirm that RAG-based methods consistently outperform CoT, a method relying on a model's intrinsic memorization and reasoning capacities. Among all RAG-based baselines, the table demonstrates that GoldRouter consistently outperforms EfficientRAG, achieving up to a 5\% improvement in GPT-evaluated accuracy. Furthermore, GoldRouter shows significant advantages over baseline methods within single-RAG-agent systems on multi-hop QA datasets (i.e., HotpotQA, 2WikiMultiHopQA, MuSiQue). This performance gain underscores the inherent challenges of managing a large-scale knowledge repository: retrieving relevant information from large-scale, aggregated knowledge bases often includes irrelevant or distracting content, consequently limiting overall performance. Such extraneous content significantly complicates multi-hop reasoning, making it challenging to effectively simplify questions by discarding resolved portions of the query. In contrast, the multi-agent approach enables each agent to concentrate on a smaller, domain-specific set of documents, enhancing the relevance and precision of information retrieval. As a result, the multi-agent system can have a higher upper bound of the final performance. 

\vspace{-5pt}
\subsection{All-Agent Participation Does NOT Work} \label{subsec:all}
\vspace{-5pt}

\begin{wraptable}{r}{0.55\textwidth}
    \vspace{-10pt}
    \centering
    \renewcommand{\arraystretch}{1.2}
    \caption{Performance comparison between GoldRouter and All-agent Participation on the NQ dataset.}
    \vspace{-8px}
\resizebox{.53\textwidth}{!}{
\begin{tabular}{l l | c c c}
    \Xhline{1.5pt}
    Methods & Models & LM & Acc. & Time \\
    \hline
    \multirow{2}{*}{\makecell[l]{All-agent\\Participation}} 
        & Qwen-Plus    
            & 50.63 & 67.76 & 72.91 \\
        & LLaMA-3.1-8B    
            & 46.19 & 56.55 & 40.18 \\
    \hline
    \multirow{2}{*}{GoldRouter} 
        & Qwen-Plus    
            & \textbf{63.41} & \textbf{83.02} & \textbf{12.97} \\
        & LLaMA-3.1-8B    
            & \underline{55.13} & \underline{69.70} & \underline{6.29} \\
    \Xhline{1.5pt}
\end{tabular}
}
    \label{tab:nq_questions}
\end{wraptable}

A naive way to avoid routing errors is \textit{all-agent participation} (Figure \ref{subfig:(problem_b)}): the server broadcasts the query to every agent and aggregates all responses. Table \ref{tab:nq_questions} shows that even on single-hop QA, this strategy is less accurate and much slower. For example, with Qwen-Plus, accuracy drops from 83.02 to 67.76 and latency rises from 12.97s to 72.91s; similar trends hold for LLaMA-3.1-8B, and the cost becomes prohibitive for multi-hop queries because broadcasting repeats across rounds. This gap is driven by irrelevant agent involvement: agents without useful evidence still consume computation, introduce distracting content, and force the server to wait for slow responses. The resulting long, noisy context makes evidence fusion harder and can reduce correctness. Therefore, a multi-RAG-agent system needs a routing mechanism that selects a small but sufficient subset of agents per round and supports knowledge fusion across rounds.

\vspace{-5pt}
\section{\algo} \label{sec: routing}
\vspace{-5pt}

\algo is designed for a practical multi-agent QA setting in which knowledge is distributed across specialized RAG agents, but users still expect a single accurate answer. In this setting, the main challenge is not only to find relevant agents, but also to combine their partial evidence into a coherent solution. A useful system therefore needs to do two things well: limit participation to agents that are likely to help, and continue reasoning when the answer must be assembled across multiple steps. 

To make this possible, \algo decomposes the system into offline and online stages, as illustrated in \cref{fig:single-hop}. At the offline stage (\cref{subfig:(routing_step_a)}), each agent periodically summarizes the structure of its local knowledge in the embedding space and shares only a few lightweight representations with the server. At the online stage (\cref{subfig:(routing_step_b)}), the server uses these summaries to route the query, judge the quality of returned responses, and iterate when necessary. In this way, routing and iterative refining-solving are handled within a single framework rather than as two disconnected procedures.
This design avoids the inefficiency of querying every agent while still supporting multi-step multi-agent evidence composition.

\vspace{-5pt}
\subsection{Offline Agent-Side Preparation: Knowledge Summarization} \label{subsec:clustering}
\vspace{-5pt}

\begin{figure*}
    \centering
    \includegraphics[width=\linewidth]{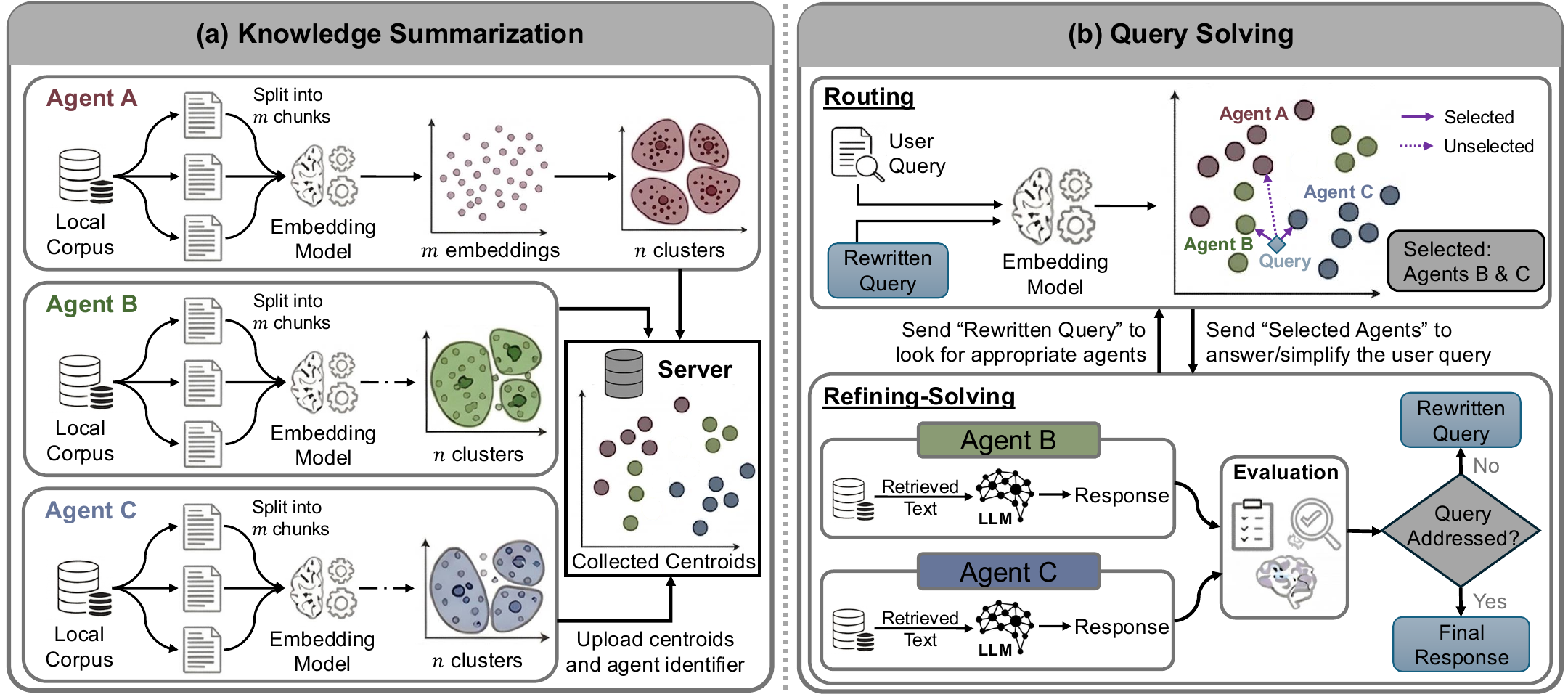}
    \vspace{-15px}
    \subfloat{\label{subfig:(routing_step_a)}}{}
    \subfloat{\label{subfig:(routing_step_b)}}{}
    \caption{Overview of \algo}
    \label{fig:single-hop}
    \vspace{-15px}
\end{figure*}

As shown in \cref{subfig:(routing_step_a)}, each agent periodically summarizes its local knowledge base into a small set of centroids for routing. Concretely, an agent splits its local corpus into $m$ chunks and encodes each chunk into an embedding vector $e_1,\dots,e_m$ using a shared embedding model, where these embeddings are cached for later retrieval process. The agent then partitions these embeddings into $n$ clusters and uploads only the resulting cluster centroids (along with an agent identifier) to the server. This design preserves knowledge sovereignty by keeping raw knowledge pieces local, while still allowing the server to estimate which RAG agents are relevant to a query in the embedding space.

To make centroids reliable routing signals, clusters must be internally coherent. If a cluster mixes semantically unrelated pieces, its centroid degenerates into an "average topic" and becomes a weak relevance signal. We therefore formulate clustering as a max-min objective that encourages \emph{every} cluster to be tight. Let $\{c_1,\dots,c_n\}$ be a partition of $\{e_1,\dots,e_m\}$ into $n$ non-empty clusters, and let $\mathrm{sim}(\cdot,\cdot)$ denote cosine similarity. We maximize the worst-case within-cluster similarity:
\begin{equation}
\label{eq:cluster-maximin}
\begin{aligned}
\{c_k^\star\}_{k=1}^n \in 
&\operatorname*{arg\,max}_{\{c_k\}_{k=1}^n}\;
\min_{k\in[n]}\ \min_{\substack{e_a,e_b\in c_k\\ a\neq b}}
\mathrm{sim}(e_a,e_b)
\; \text{s.t.}\;
\{c_k\}_{k=1}^n \text{ is a partition of } \{e_1,\dots,e_m\}.
\end{aligned}
\end{equation}
This objective prevents the optimization from sacrificing low-quality clusters because the least coherent cluster controls the objective value. As a result, each centroid is a faithful representative of a semantically consistent region of the agent's knowledge base. 

In practice, each agent optimizes \cref{eq:cluster-maximin} through an approximate max-min similarity clustering algorithm (\cref{algo}) to obtain $n$ disjoint clusters over its local embeddings. 
Starting from singleton clusters, the algorithm first normalizes the embeddings, computes pairwise cosine similarities, and iteratively merges the pair of current clusters with the largest bottleneck similarity until only $n$ clusters remain. It then performs a local refinement stage that identifies the cluster with the smallest within-cluster bottleneck similarity, considers reassigning one endpoint of its worst-similarity pair to another cluster, and accepts the move only if it improves the objective in \cref{eq:cluster-maximin}. After obtaining the final partition, the agent computes one centroid for each cluster, i.e., the mean of its member embeddings, and uploads only these centroids to the server.
In our experiments, we set $n = \lfloor \sqrt{m} \rfloor$, where $m$ is the number of local knowledge chunks. Full algorithmic details are provided in Appendix \ref{apdx:opt_algo}, with pseudocode in \cref{algo}.

\vspace{-5pt}
\subsubsection{Analysis of the Effectiveness}
\vspace{-5pt}

\begin{theorem}[Query-to-Agent Routing Bound]
Let $u$ be a query embedding, and let agent $\alpha$ contain a knowledge-piece embedding
$e$ such that $\operatorname{sim}(u,e)\ge k$ for some $k\in(0,1]$.
Assume $e$ belongs to an $s$-coherent cluster $c$ of agent $\alpha$, meaning
$\operatorname{sim}(e',e'')\ge s$ for all $e',e''\in c$, with centroid $\bar c$.
Let
\begin{equation}
R_{\alpha}(u):=\max_{\bar c' \in C_{\alpha}}\operatorname{sim}(u,\bar c')
\end{equation}
be the routing score of agent $\alpha$.
Then
\begin{equation}
R_{\alpha}(u)\;\ge\;\operatorname{sim}(u,\bar c)
\;\ge\;
ks-\sqrt{(1-k^2)(1-s^2)}.
\end{equation}
Equivalently, the routing surrogate incurs at most
\begin{equation}
\Delta_{\alpha}(u;e):=
\big[\operatorname{sim}(u,e)-R_{\alpha}(u)\big]_+
\;\le\;
k-ks+\sqrt{(1-k^2)(1-s^2)}.
\end{equation}
\end{theorem}

The detailed proof is provided in Appendix \ref{apdx:clustering}. The theorem shows that centroid routing is a stable surrogate for fine-grained matching. If a query is close to a knowledge piece inside an agent, and that piece lies in a coherent cluster, then the agent's centroid score cannot drop arbitrarily. Hence, the server can route using only uploaded centroids while incurring a bounded loss relative to routing with the underlying knowledge pieces. 

\vspace{-5pt}
\subsection{Online Query Solving via Routing and Iterative Refining-Solving} \label{subsec:online}
\vspace{-5pt}

As shown in \cref{subfig:(routing_step_b)}, the uploaded centroids provide the server with an embedding-space approximation of each agent's \emph{knowledge capability}: each centroid represents a coherent region where the agent is likely to contain supporting evidence. Given a query, the server embeds it and compares the query embedding with the uploaded centroids to identify the most appropriate agents. After receiving responses from the selected agents, the server evaluates whether those responses adequately address the query. If the query is not fully resolved, the server rewrites it to focus on the remaining unresolved aspects and then repeats the routing and response-evaluation process until a final answer is produced or no further progress can be made.

\NewDocumentCommand{\rtans}{g}{%
  \IfNoValueTF{#1}
    {\ensuremath{\operatorname{\textsc{RtAns}}}\xspace}%
    {\ensuremath{\operatorname{\textsc{RtAns}}(#1)}}%
}

\vspace{-5pt}
\paragraph{Single-round query solving (\rtans).}
Given a query $q$, the server embeds it as $x=\phi(q)$ using the same embedding model $\phi(\cdot)$ used by agents. Let $\mathcal{C}$ be the set of all centroids from all RAG agents, and let $f(\bar c)$ map a centroid $\bar c$ to its source agent. The server selects the Top$K$ centroids by similarity and routes the query to the corresponding agents:
\begin{equation}
\label{eq:topk-routing}
\scalebox{1.0}{$\displaystyle
\mathcal{C}_K(q)=\operatorname{TopK}_{\bar c\in\mathcal{C}}\mathrm{sim}(x,\bar c),
\;
\mathcal{A}(q)=\{f(\bar c):\bar c\in\mathcal{C}_K(q)\}
$}
\end{equation}
Each selected RAG agent answers the query using its local knowledge and returns an evidence-backed response. The server then applies \emph{Response Evaluator} to label each response as \textit{addressed} vs.\ \textit{not addressed} and curates the round output (e.g., selecting or combining \textit{addressed} responses). We denote this single-round procedure as \rtans{q}, which includes: (i) Top$K$ centroid-based agent selection, (ii) agent response generation with local evidence, (iii) server-side response evaluation, and (iv) response curation. For single-hop queries, a single \rtans call is typically sufficient.

\vspace{-5pt}
\paragraph{Iterative refining-solving for multi-hop queries.}
For multi-hop queries, the required evidence may be distributed across agents and across reasoning steps, so a single \rtans{q} call can be insufficient. \algo repeats \rtans in an iterative loop. At iteration $t$, the server calls \rtans{q_t} and uses the Response Evaluator to label responses as fully addressed, partially addressed, or not addressed. If any response is fully addressed, the Response Summarizer consolidates the evidence-backed responses into the final answer and terminates. If only partially addressed responses exist, the server aggregates partial progress, invokes the Question Simplifier to produce a refined query $q_{t+1}$ targeting unresolved aspects, and then proceeds to the next iteration. If all responses are not addressed, the server stops and reports that the query is unanswerable within the system scope.
Full step-by-step procedures and server-module prompts are provided in Appendix~\ref{apdx:rirs_more}. 

\vspace{-5pt}
\paragraph{Practical properties.} 
\algo has three practical advantages. First, centroid-based routing activates only a small set of relevant agents for each query, reducing fan-out, token usage, and interference from irrelevant responses. Second, for complex questions, the iterative refining-solving loop turns partial progress into targeted follow-up queries, enabling evidence composition across agents and across rounds. Third, RIRS cleanly separates offline knowledge summarization from online query solving: agents can update local knowledge, refresh their centroids, and synchronize them with the server, while the server maintains only a lightweight routing index. This separation simplifies new agent onboarding, supports continuous knowledge updates, and avoids centralizing raw corpora or retraining a router.

\vspace{-5pt}
\section{Experiments}
\vspace{-5pt}

\subsection{Experimental Setup} \label{subsec:setup}
\vspace{-5pt}

Due to space limits, this section summarizes the experimental setup needed to interpret our quantitative results. Full implementation details and baseline specifications are provided in Appendix~\ref{apdx:baselines}. 

\vspace{-5px}
\paragraph{Datasets.} 
Our experiments cover both single-hop and multi-hop open-ended QA tasks within a unified evaluation framework. We employ the Natural Questions (NQ) dataset \cite{kwiatkowski2019natural} for single-hop QA to emphasize the effectiveness of the routing mechanism. For evaluating the performance on answering complex questions, we assess performance on four multi-hop QA benchmarks: HotpotQA \cite{yang2018hotpotqa}, 2WikiMultiQA \cite{ho2020constructing}, MusiQue \cite{trivedi2022musique}, and Multi-Hop RAG (MHR) Benchmark \cite{tang2024multihop}. Except for MHR, all questions in these datasets are answerable using knowledge from Wikipedia.

\vspace{-5px}
\paragraph{Models.}
We use \texttt{text-embedding-v2}~\cite{zhang2024mgte} to compute dense representations for retrieval.
Agents are powered by two LLM backends: \texttt{llama-3.1-8b-instruct}~\cite{touvron2023llama,touvron2023llama2,dubey2024llama} and \texttt{qwen-plus-2024-12-20}~\cite{yang2024qwen2}.
All methods are evaluated in an off-the-shelf setting without fine-tuning.

\vspace{-5px}
\paragraph{Multi-RAG-Agent Systems.} 
Throughout the experiments, we construct two multi-RAG-agent systems tailored to different datasets. 
\begin{itemize}[topsep=0pt,itemsep=0pt,parsep=0pt,partopsep=0pt,leftmargin=*]
    \item \textbf{WikiAgents for NQ, HotpotQA, 2WikiMultiQA and MusiQue:} WikiAgents group is built upon a corpus of over 121K Wikipedia pages, dumped as of November 1, 2023, and made publicly available via the HuggingFace dataset. The system comprises exactly 64 RAG agents. This specific number is derived from the inherent limitation of the \href{https://ores.wikimedia.org/docs}{ORES legacy service}, which can only classify a Wikipedia page into 64 predefined categories \cite{johnson2021classification}. Consequently, each RAG agent is designated to handle one of these 64 categories, ensuring that the categorization of pages is consistent and aligned with the predefined taxonomy established by the ORES service. However, these agents cannot cover all required documents, and once the question cannot be answered, a knowledgeable agent will be called to answer the question because the existing LLMs have been pretrained with the Wikipedia corpus.
    \item \textbf{NewsAgents for MHR:} NewsAgents group is constructed on a corpus of 609 news articles from 49 distinct news media and spanning six domains \cite{tang2024multihop}. To analyze the effects of source and domain characteristics, we further split NewsAgent into two variants: \textit{NewsAgent-Source} and \textit{NewsAgent-Domain}, corresponding to partitions based on news media and domain categories, respectively. Each agent uses LLaMAIndex to split news articles into 256-token chunks, with 20 overlapping tokens between consecutive chunks. 
\end{itemize}

\vspace{-5px}
\paragraph{Baselines.} 
We compare \algo with two representative multi-agent routing baselines. \emph{Chameleon} \cite{lu2024chameleon} is a description-based router that maintains a textual description for each agent and prompts an LLM to select the most relevant agents for a given query. \emph{RouterDC} \cite{chen2024routerdc} is a learning-based router that constructs agent representations from single-hop answerable questions and routes by embedding similarity while keeping the embedding model fixed. Both methods operate in the same multi-RAG-agent setting as \algo and serve as alternative routing strategies. Further implementation details are provided in Appendix \ref{sec:baselines}.

\vspace{-5px}
\paragraph{Evaluation Metrics.}
We report three metrics.
\emph{Lexical Match (LM)} measures the percentage of questions whose ground-truth answer string appears in the generated response.
\emph{GPT Evaluation (Acc.)} uses \texttt{gpt-4o-2024-08-06}~\cite{achiam2023gpt} to judge correctness by comparing model outputs against ground-truth answers while allowing paraphrases.
\emph{Time} measures wall-clock latency from receiving the question to producing a seemingly valid response; this metric can be noisy due to network/API variance and runtime contention.

\vspace{-5px}
\subsection{Ablation Studies}
\vspace{-5pt}

\begin{table*}[t]
  \centering
  \renewcommand{\arraystretch}{1.05}
  \caption{Performance under different numbers of selected clusters on the Natural Questions dataset}
  \vspace{-10px}
  \resizebox{\linewidth}{!}{
  \begin{tabular}{cccccccccc}
    \Xhline{1.5pt}
    \multirow{2}{*}{\makecell[c]{\# Sel.\\Clusters}} & \multirow{2}{*}{\makecell[c]{Avg Sel. Agents\\(Eff. Rate)}} & \multirow{2}{*}{\makecell[c]{Agent-Level\\Ans. Rate}} & \multirow{2}{*}{\makecell[c]{Avg Viewed Doc.\\(Eff. Rate)}} & \multirow{2}{*}{\makecell[c]{Doc.-Level\\Ans. Rate}} & \multicolumn{2}{c}{LLaMA-3.1-8B} && \multicolumn{2}{c}{Qwen-Plus} \\\cline{6-7} \cline{9-10}
    & & & & &  LM & Acc. && LM & Acc. \\\hline
    1  & 1.0 (52.69\%)   & 52.69\%   & 3.32 (21.43\%)   & 50.58\%   & 43.53\%   & 60.91\%   && 59.90\%   & 77.72\%   \\ 
    5  & 3.32 (37.47\%)  & 76.47\%   & 10.12 (11.61\%)  & 73.01\%   & 50.98\%   & 63.96\%   && 62.05\%   & 80.05\%   \\ 
    10 & 5.28 (26.76\%)  & 84.52\%   & 16.95 (7.07\%)   & 80.12\%   & 50.18\%   & 64.74\%   && 62.90\%   & 81.05\%   \\ \Xhline{1.5pt}
  \end{tabular}
  }
  \vspace{-10px}
  \label{tab:topk_selection}
\end{table*}

\paragraph{Choice of TopK.} 
Table~\ref{tab:topk_selection} shows that selecting more clusters increases both agent-level and document-level answerable rates, which generally improves end-to-end LM/Acc.\ at the cost of invoking more agents and viewing more documents. 
A key observation is that increasing TopK from 1 to 5 yields a substantial gain, while further increasing it from 5 to 10 brings only marginal improvement, suggesting diminishing returns beyond moderate agent coverage. Notably, the final accuracy can exceed the answerable rate because the system falls back to a CoT-only response when the selected agents are judged unable to answer; in such cases, the LLM may still answer correctly using parametric knowledge. Finally, document-level answerable rates are consistently lower than agent-level rates, indicating that even when the \emph{right} agent is selected, retrieval within its local index may still miss the most relevant evidence.

\vspace{-5px}
\subsection{Analysis with Wikipedia-related QA} \label{subsec: single-hop-exp}
\vspace{-5pt}

\begin{table*}[t]
    \centering
    \caption{Performance comparison of different methods under various datasets and the knowledge of WikiAgents. The best results under Qwen-Plus and LLaMA-3.1-8B are \textbf{Bold} and \underline{underline} out, respectively, among the methods to multiple RAG agents (i.e., Chameleon, RouterDC, and \algo), which are highlighted in the \colorbox{lightgray}{grey} background.}
    \vspace{-10px}
    \definecolor{lightgray}{gray}{0.9}
    \renewcommand{\arraystretch}{1.3}
    \resizebox{\textwidth}{!}{
    \begin{threeparttable}
    \begin{tabular}{l l | c c c | c c c c c | c c c c c | c c c c c }
        \Xhline{2pt}
        \multirow{2}{*}{Methods} & \multirow{2}{*}{Models} & \multicolumn{3}{c|}{Natural Questions} & \multicolumn{5}{c|}{HotpotQA} & \multicolumn{5}{c|}{2WikiMultiHopQA} & \multicolumn{5}{c}{MuSiQue} \\\cline{3-20}
        & & LM & Acc. & Time & LM & F1 & Acc. & Rounds & Time & LM & F1 & Acc. & Rounds & Time & LM & F1 & Acc. & Rounds & Time  \\
        \hline
        \multirow{2}{*}{GoldRouter} 
            & Qwen    
                & 63.41 & 83.02 & 12.97 
                & 58.17 & 63.99 & 81.01 & 1.22 & 22.85 
                & 70.11 & 72.37 & 76.33 & 1.36 & 30.93 
                & 29.01 & 33.75 & 47.55 & 1.51 & 29.87 \\
            & LLaMA    
                & 55.13 & 69.70 & 6.29 
                & 50.60 & 56.26 & 64.94 & 1.55 & 17.07 
                & 49.73 & 52.99 & 52.07 & 1.58 & 20.57 
                & 23.27 & 28.18 & 31.46 & 1.86 & 21.68 \\\hline
        \rowcolor{lightgray} 
            & Qwen    
                & 55.48 & 73.92 & 24.90 
                & 52.12 & 57.55 & 73.19 & 1.43 & 39.05 
                & 56.68 & 59.18 & 60.12 & 1.70 & 37.53 
                & \textbf{24.21} & \textbf{28.88} & 39.87 & 1.77 & 48.46 \\
        \rowcolor{lightgray}
        \multirow{-2}{*}{Chameleon}
            & LLaMA    
                & 49.68 & 62.73 & 15.27 
                & 41.08 & 46.59 & 50.20 & 1.85 & 24.77 
                & 40.52 & 43.49 & 40.27 & 1.84 & 26.71 
                & 15.69 & 20.11 & 23.74 & 2.08 & 32.81 \\\hline
        \rowcolor{lightgray} 
            & Qwen    
                & 53.93 & 71.09 & 24.91 
                & 51.34 & 56.84 & 67.24 & 1.46 & 44.12 
                & 50.97 & 53.88 & 52.12 & 1.67 & 45.61 
                & 23.38 & 27.69 & 37.43 & 1.71 & 52.98 \\
        \rowcolor{lightgray}
        \multirow{-2}{*}{RouterDC}
            & LLaMA    
                & 47.26 & 60.43 & 13.11 
                & 38.31 & 43.11 & 47.74 & 1.78 & 23.63 
                & 34.48 & 37.35 & 30.59 & 1.67 & 26.58 
                & 15.08 & 19.11 & 21.58 & 1.98 & 29.66 \\\hline
        \rowcolor{lightgray} 
            & Qwen    
                & \textbf{62.05} & \textbf{80.05} & 23.30 
                & \textbf{54.01} & \textbf{59.62} & \textbf{75.63} & 1.44 & 34.71 
                & \textbf{61.14} & \textbf{63.71} & \textbf{65.78} & 1.73 & 39.76 
                & 23.81 & 28.63 & \textbf{41.24} & 1.67 & 43.98 \\
        \rowcolor{lightgray}
            & LLaMA    
                & \underline{50.98} & \underline{63.96} & 10.51 
                & \underline{47.57} & \underline{52.56} & \underline{58.69} & 2.04 & 19.93 
                & \underline{45.65} & \underline{49.10} & \underline{45.65} & 2.02 & 22.26 
                & \underline{20.16} & \underline{24.22} & \underline{25.82} & 2.48 & 25.14 \\
        \rowcolor{lightgray}
        \multirow{-3}{*}{\algo}
            & Mixed\tnote{1}   
                & 55.98 & 73.32 & 20.98 
                & 51.68 & 56.65 & 69.33 & 1.83 & 41.04 
                & 59.48 & 61.94 & 62.06 & 1.85 & 39.37 
                & 24.50 & 28.57 & 39.75 & 2.08 & 47.51 \\
        \Xhline{2pt}
    \end{tabular}
    \begin{tablenotes}
      \item[1] The RAG agents use LLaMA-3.1-8B, while other modules in the server use Qwen-Plus.
    \end{tablenotes}
    \end{threeparttable}}
    \vspace{-17px}
    \label{tab:wiki_comparison}
\end{table*}

\vspace{-3px}
Table~\ref{tab:wiki_comparison} compares the proposed \algo with representative multi-agent routing baselines on four Wikipedia-based QA benchmarks under the WikiAgents corpus. We summarize the most salient findings below. 

\vspace{-5px}
\paragraph{Comparison with GoldRouter.} 
As an error-free routing mechanism, GoldRouter serves as the upper bound in a multi-agent system, achieving the highest scores in LM, F1, and Accuracy, as well as the lowest values in rounds and response time. In contrast, all other methods (Chameleon, RouterDC, and \algo) experience varying degrees of performance degradation. i.e., more query rounds but with lower accuracy. The degradation is owed to routing errors, highlighting the importance of designing an effective routing mechanism and underscoring the inherent challenge of dynamically estimating each agent's expertise on-the-fly.

\vspace{-5px}
\paragraph{Comparison with other routing methods.} 
From Table~\ref{tab:wiki_comparison}, the accuracy ranking among the three routing methods consistently follows: \algo $>$ Chameleon $>$ RouterDC, under the same dataset and backbone LLM. The advantage of \algo is especially clear on harder multi-hop benchmarks and with the weaker backbone (LLaMA-3.1-8B), where \algo improves accuracy by up to 8\% over Chameleon and 15\% over RouterDC. 
These results suggest that \algo more reliably identifies agent knowledge boundaries. Concretely, \algo builds cluster-level representations to characterize each agent's knowledge capacity, enabling more accurate routing. By contrast, RouterDC approximates agent expertise via a fixed cache of historically answerable questions, and Chameleon relies on static textual descriptions; both signals can be incomplete and lead to misrouting. Notably, while \algo may use slightly more rounds than Chameleon, it often runs faster overall because it avoids additional LLM calls during agent selection. 

\vspace{-5px}
\paragraph{Comparison with various LLMs.} 
Using a stronger backbone (Qwen-Plus) consistently improves all methods over LLaMA-3.1-8B, as expected. Moreover, the mixed setting, which uses a stronger server-side controller (Qwen-Plus) with lightweight agent models, recovers much of the performance of Qwen-Plus while keeping agent-side costs low. Together with the response-evaluator results in Table~\ref{tab:performance}, this suggests that a strong evaluator/controller is critical for reducing error propagation when intermediate agent responses are noisy. This design makes \algo practical for deployment: Lightweight models can run at the RAG agents, allowing the underlying knowledge bases to remain local and generating evidence-backed responses even in resource-constrained settings. Meanwhile, the user-facing server is equipped with a stronger model to ensure the final response is precise.

\vspace{-5px}
\subsection{Analysis for Multi-hop RAG}
\vspace{-5pt}

\begin{wrapfigure}{r}{0.6\textwidth}
    \begin{minipage}{0.6\textwidth}
        \centering

        \vspace{-5pt}
        \scriptsize
        \setlength{\tabcolsep}{3.2pt}
        \renewcommand{\arraystretch}{1.2}

        \begin{threeparttable}
        \begin{tabularx}{\linewidth}{X l | c c c | c c c }
            \Xhline{1.5pt}
            \multirow{2}{*}{Methods} & \multirow{2}{*}{Models} & \multicolumn{3}{c|}{NewsAgent-Source} & \multicolumn{3}{c}{NewsAgent-Domain} \\\cline{3-8}
            &  & Acc. & Hall.\tnote{1} & Rounds & Acc. & Hall.\tnote{1} & Rounds \\
            \hline
            \multicolumn{8}{l}{\textit{Single-RAG-Agent System}} \\\hline
            \multirow{2}{*}{EfficientRAG} & Qwen & 81.80 & 1.19 & 1.57 & - & - & - \\
            & LLaMA & 66.80 & 6.05 & 1.32 & - & - &-  \\\hline
            \multicolumn{8}{l}{\textit{Multi-RAG-Agent System}} \\\hline
            \multirow{2}{*}{GoldRouter} & Qwen & 89.32 & - & 1.06 & 90.31 & - & 1.12 \\
            & LLaMA & 79.53 & - & 1.08 & 80.16 & - & 1.18 \\\hline
            \multirow{2}{*}{Chameleon} & Qwen & 76.33 & 5.76 & 1.84 & 86.24 & 1.15 & 1.73 \\
            & LLaMA & 54.52 & 11.37 & 1.31 & 72.91 & 5.20 & 1.44 \\\hline
            \multirow{2}{*}{\algo} & Qwen & 84.75 & 0.93 & 1.65 & 88.78 & 1.07 & 1.66 \\
            & LLaMA & 71.91 & 4.91 & 1.37 & 75.12 & 4.73 & 1.39 \\
            \Xhline{1.5pt}
        \end{tabularx}%
        \begin{tablenotes}
          \item[1] Hall. means a hallucination rate, indicating the proportion of null queries incorrectly judged as answerable among all queries classified as answerable.
        \end{tablenotes}
        \end{threeparttable}
        \vspace{-5pt}
        \captionof{table}{Performance comparison under NewsAgent-Source and NewsAgent-Domain. Single-agent results are reported only for NewsAgent-Source because both settings use the same knowledge repository in that case.}
        \label{tab:newsagent_comparison}

        \vspace{-3pt}

        \normalsize
        \subfloat[Qwen-Plus]{%
            \includegraphics[width=0.48\linewidth]{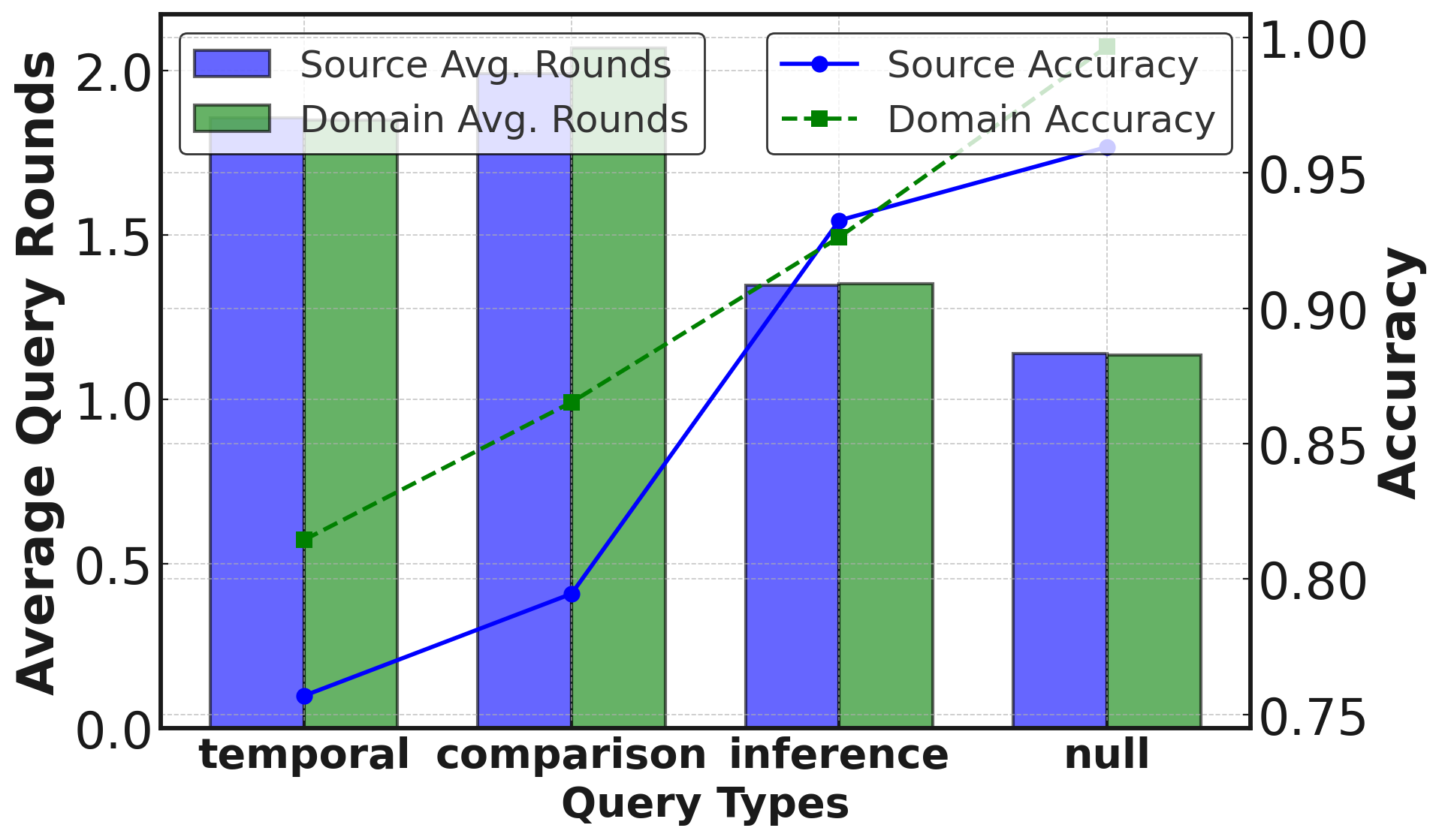}
            \label{fig:mhr_qwen}
        }
        \subfloat[LLaMA-3.1-8B]{%
            \includegraphics[width=0.48\linewidth]{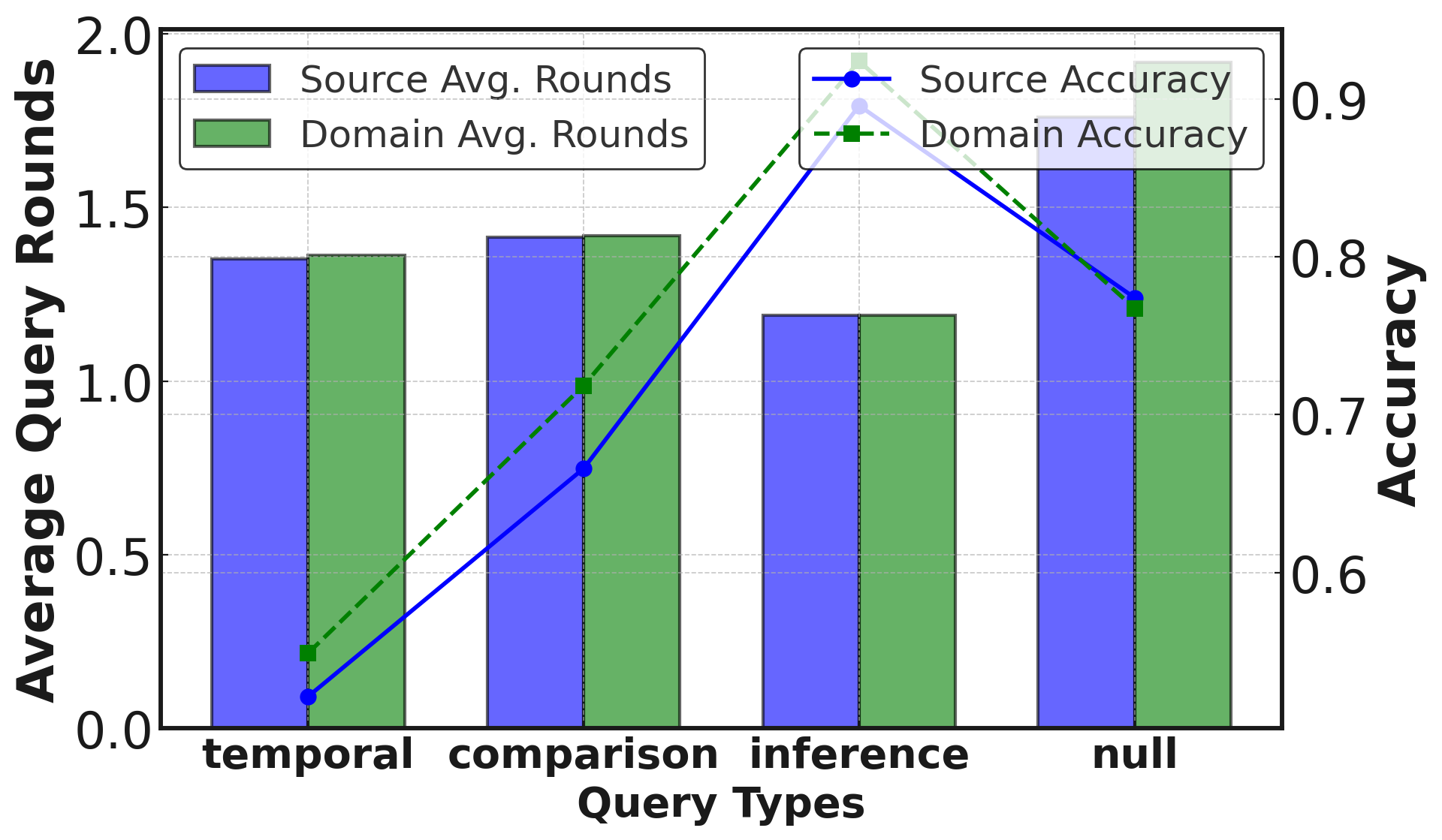}
            \label{fig:mhr_llama}
        }
        \vspace{-5pt}
        \captionof{figure}{Performance comparison on MHR under different query types.}
        \label{fig:mhr}
    \end{minipage}
    \vspace{-10pt}
\end{wrapfigure}

This section analyzes the performance of our proposed method, \algo, on the MHR benchmark under two NewsAgents settings: \textbf{NewsAgent-Source} (49 agents) and \textbf{NewsAgent-Domain} (six agents). As shown in Table~\ref{tab:newsagent_comparison} and Figure~\ref{fig:mhr}, we focus on two aspects: (i) comparing \algo with single-agent RAG baselines, and (ii) examining how the number of RAG agents affects performance.

\vspace{-5pt}
\paragraph{Comparison with a Single-RAG-Agent System.} 
\cref{tab:newsagent_comparison} compares \algo with both single-agent and multi-agent baselines under the NewsAgent-Source and NewsAgent-Domain settings.
In the single-agent setting, all documents are pooled into one centralized RAG system and answered with EfficientRAG. Although this removes routing, it also enlarges the retrieval space and increases the chance of introducing irrelevant evidence. By contrast, \algo partitions knowledge among specialized agents and iteratively routes each query to the most relevant agents, thereby reducing retrieval noise and cross-topic interference. As shown in \cref{tab:newsagent_comparison}, \algo achieves higher accuracy and lower hallucination rates than the single-agent baselines. The advantage is especially clear for multi-hop questions, where source-based or domain-based partitions help isolate the evidence needed for each reasoning step. Compared with Chameleon, the gap is particularly pronounced under NewsAgent-Source, where outlet-level descriptions are often too coarse because a single outlet can cover many unrelated topics. In the more structured NewsAgent-Domain setting, Chameleon becomes stronger but still lags behind \algo.

\vspace{-8px}
\paragraph{Discussion of Different Numbers of RAG Agents.} 
From Table~\ref{tab:newsagent_comparison} and Figure~\ref{fig:mhr}, we observe a trade-off between granularity and routing reliability. Using more agents provides finer-grained coverage and typically reduces the number of rounds needed to locate relevant evidence. However, a larger agent pool also increases routing uncertainty and can hurt accuracy. Compared with GoldRouter, performance degradation is more pronounced under NewsAgent-Source than under NewsAgent-Domain. We attribute this gap to suboptimal agent formation in NewsAgent-Source: partitioning by news outlets yields agents with diverse, weakly structured topic coverage, which makes it harder to route queries to the right agents. Besides, the much larger number of agents in NewsAgent-Source further amplifies routing uncertainty.

A closer comparison between the single-RAG-agent system and NewsAgent-Domain reveals the existence of an \textit{optimal number} of agents for a given knowledge base. Using too few agents can lead to overly broad domain coverage, which reintroduces the challenge of retrieving irrelevant information. In contrast, the domain-based configuration results in more efficient and accurate resolution of complex, multi-hop questions.

\vspace{-8pt}
\section{Conclusion}
\vspace{-8pt}

We study how to reliably answer user queries when knowledge is distributed across multiple business units and cannot be centralized due to privacy, compliance, or ownership constraints. To this end, we propose \algo, an iterative routing framework that coordinates multiple RAG agents while preserving knowledge sovereignty. \algo decouples \emph{offline} knowledge summarization from \emph{online} query solving: each agent summarizes its local corpus into centroid-based representations in the embedding space and maintains these summaries independently, while a user-facing server performs iterative refining-solving and fuses evidence across agents and rounds. Experiments show that \algo consistently outperforms strong multi-agent baselines, and its architecture supports practical deployment: agents can run lightweight models for grounded generation with local corpora, while the server uses a stronger model to synthesize reliable responses. 

\section*{Ethics Statement}

This work studies a realistic multi-agent question-answering setting in which knowledge is distributed across RAG agents and cannot be centralized. The proposed framework reduces unnecessary data exposure by sharing embedding-space summaries for routing. However, this design does not address two potential risks. In an \emph{agent-in-the-wild} setting, malicious agents may return incomplete, misleading, or low-quality responses, and routing or evaluation errors may propagate across iterations, especially for complex multi-hop questions. To reduce this risk, RIRS requires evidence-backed responses, filters returned outputs through a server-side evaluator, and allows the system to stop or declare a query unanswerable when adequate support is unavailable. A second concern is privacy leakage: Similar to federated learning, the proposed \algo offers a fundamental privacy protection because centroid summaries are more restrictive than sharing raw documents. However, it still reveals partial information about local knowledge, especially under repeated probing or after updates to an agent's corpus. An important direction for future work is to study how additional privacy-preserving mechanisms, such as Gaussian or Laplace perturbation, affect routing accuracy, and to develop stronger privacy-preserving summarization and access-control mechanisms that preserve comparable performance. 

\section*{Reproducibility Statement}

We make reproducibility a priority in both the algorithmic design and the experimental setup. The paper specifies the full \algo pipeline, including the offline knowledge summarization stage, the online query solving, the clustering objective, and the corresponding approximation algorithm. The appendix further provides pseudocode, theoretical analysis, step-by-step procedures, and the prompt templates used by the server modules and the RAG agents. The experimental section reports the datasets, corpus construction, backbone models, baselines, evaluation metrics, and implementation details needed to replicate the reported results. All methods are evaluated in an off-the-shelf setting without task-specific fine-tuning. At the same time, exact reproduction remains challenging because LLM-based generation is inherently stochastic, and this variance is amplified in multi-hop settings where later rounds depend on earlier query rewriting and response evaluation. To reduce the impact of randomness, each experiment is run three times, and the paper reports average results. We also disclose the main system components used in our implementation, including retrieval, chunking, model serving, and agent orchestration, and have released our code repository at \href{https://github.com/HarliWu/RIRS}{GitHub} to facilitate close reproduction of the experiments.


\bibliography{custom}

\begin{thebibliography}{88}
\providecommand{\natexlab}[1]{#1}
\providecommand{\url}[1]{\texttt{#1}}
\expandafter\ifx\csname urlstyle\endcsname\relax
  \providecommand{\doi}[1]{doi: #1}\else
  \providecommand{\doi}{doi: \begingroup \urlstyle{rm}\Url}\fi

\bibitem[Achiam et~al.(2023)Achiam, Adler, Agarwal, Ahmad, Akkaya, Aleman, Almeida, Altenschmidt, Altman, Anadkat, et~al.]{achiam2023gpt}
Josh Achiam, Steven Adler, Sandhini Agarwal, Lama Ahmad, Ilge Akkaya, Florencia~Leoni Aleman, Diogo Almeida, Janko Altenschmidt, Sam Altman, Shyamal Anadkat, et~al.
\newblock Gpt-4 technical report.
\newblock \emph{arXiv preprint arXiv:2303.08774}, 2023.

\bibitem[Addison et~al.(2024)Addison, Nguyen, Medan, Manzari, McElrone, Lalwani, More, Sharma, Roth, Yang, et~al.]{addison2024c}
Parker Addison, Minh-Tuan~H Nguyen, Tomislav Medan, Mohammad~T Manzari, Brendan McElrone, Laksh Lalwani, Aboli More, Smita Sharma, Holger~R Roth, Isaac Yang, et~al.
\newblock C-fedrag: A confidential federated retrieval-augmented generation system.
\newblock \emph{arXiv preprint arXiv:2412.13163}, 2024.

\bibitem[Asai et~al.(2023{\natexlab{a}})Asai, Min, Zhong, and Chen]{asai2023retrieval}
Akari Asai, Sewon Min, Zexuan Zhong, and Danqi Chen.
\newblock Retrieval-based language models and applications.
\newblock In \emph{Proceedings of the 61st Annual Meeting of the Association for Computational Linguistics (Volume 6: Tutorial Abstracts)}, pp.\  41--46, 2023{\natexlab{a}}.

\bibitem[Asai et~al.(2023{\natexlab{b}})Asai, Wu, Wang, Sil, and Hajishirzi]{asai2023self}
Akari Asai, Zeqiu Wu, Yizhong Wang, Avirup Sil, and Hannaneh Hajishirzi.
\newblock Self-rag: Learning to retrieve, generate, and critique through self-reflection.
\newblock \emph{arXiv preprint arXiv:2310.11511}, 2023{\natexlab{b}}.

\bibitem[Bai et~al.(2023)Bai, Bai, Chu, Cui, Dang, Deng, Fan, Ge, Han, Huang, et~al.]{bai2023qwen}
Jinze Bai, Shuai Bai, Yunfei Chu, Zeyu Cui, Kai Dang, Xiaodong Deng, Yang Fan, Wenbin Ge, Yu~Han, Fei Huang, et~al.
\newblock Qwen technical report.
\newblock \emph{arXiv preprint arXiv:2309.16609}, 2023.

\bibitem[Borgeaud et~al.(2022)Borgeaud, Mensch, Hoffmann, Cai, Rutherford, Millican, Van Den~Driessche, Lespiau, Damoc, Clark, et~al.]{borgeaud2022improving}
Sebastian Borgeaud, Arthur Mensch, Jordan Hoffmann, Trevor Cai, Eliza Rutherford, Katie Millican, George~Bm Van Den~Driessche, Jean-Baptiste Lespiau, Bogdan Damoc, Aidan Clark, et~al.
\newblock Improving language models by retrieving from trillions of tokens.
\newblock In \emph{International conference on machine learning}, pp.\  2206--2240. PMLR, 2022.

\bibitem[Chan et~al.(2024)Chan, Xu, Yuan, Luo, Xue, Guo, and Fu]{chan2024rq}
Chi-Min Chan, Chunpu Xu, Ruibin Yuan, Hongyin Luo, Wei Xue, Yike Guo, and Jie Fu.
\newblock Rq-rag: Learning to refine queries for retrieval augmented generation.
\newblock \emph{arXiv preprint arXiv:2404.00610}, 2024.

\bibitem[Chen et~al.(2017)Chen, Fisch, Weston, and Bordes]{chen2017reading}
Danqi Chen, Adam Fisch, Jason Weston, and Antoine Bordes.
\newblock Reading {W}ikipedia to answer open-domain questions.
\newblock In \emph{Proceedings of the 55th Annual Meeting of the Association for Computational Linguistics (Volume 1: Long Papers)}, pp.\  1870--1879, 2017.

\bibitem[Chen et~al.(2023{\natexlab{a}})Chen, Dong, Shu, Zhang, Sesay, Karlsson, Fu, and Shi]{chen2023autoagents}
Guangyao Chen, Siwei Dong, Yu~Shu, Ge~Zhang, Jaward Sesay, B{\"o}rje~F Karlsson, Jie Fu, and Yemin Shi.
\newblock Autoagents: A framework for automatic agent generation.
\newblock \emph{arXiv preprint arXiv:2309.17288}, 2023{\natexlab{a}}.

\bibitem[Chen et~al.(2024{\natexlab{a}})Chen, Jiang, Lin, Kwok, and Zhang]{chen2024routerdc}
Shuhao Chen, Weisen Jiang, Baijiong Lin, James~T Kwok, and Yu~Zhang.
\newblock Routerdc: Query-based router by dual contrastive learning for assembling large language models.
\newblock \emph{arXiv preprint arXiv:2409.19886}, 2024{\natexlab{a}}.

\bibitem[Chen et~al.(2023{\natexlab{b}})Chen, Su, Zuo, Yang, Yuan, Chan, Yu, Lu, Hung, Qian, et~al.]{chen2023agentverse}
Weize Chen, Yusheng Su, Jingwei Zuo, Cheng Yang, Chenfei Yuan, Chi-Min Chan, Heyang Yu, Yaxi Lu, Yi-Hsin Hung, Chen Qian, et~al.
\newblock Agentverse: Facilitating multi-agent collaboration and exploring emergent behaviors.
\newblock In \emph{The Twelfth International Conference on Learning Representations}, 2023{\natexlab{b}}.

\bibitem[Chen et~al.(2024{\natexlab{b}})Chen, Liu, Wang, Liu, Zhang, Chen, and Zhao]{chen2024mindsearch}
Zehui Chen, Kuikun Liu, Qiuchen Wang, Jiangning Liu, Wenwei Zhang, Kai Chen, and Feng Zhao.
\newblock Mindsearch: Mimicking human minds elicits deep ai searcher.
\newblock \emph{arXiv preprint arXiv:2407.20183}, 2024{\natexlab{b}}.

\bibitem[Du et~al.(2024)Du, Li, Torralba, Tenenbaum, and Mordatch]{du2024improving}
Yilun Du, Shuang Li, Antonio Torralba, Joshua~B Tenenbaum, and Igor Mordatch.
\newblock Improving factuality and reasoning in language models through multiagent debate.
\newblock In \emph{Forty-first international conference on machine learning}, 2024.

\bibitem[Dubey et~al.(2024)Dubey, Jauhri, Pandey, Kadian, Al-Dahle, Letman, Mathur, Schelten, Yang, Fan, et~al.]{dubey2024llama}
Abhimanyu Dubey, Abhinav Jauhri, Abhinav Pandey, Abhishek Kadian, Ahmad Al-Dahle, Aiesha Letman, Akhil Mathur, Alan Schelten, Amy Yang, Angela Fan, et~al.
\newblock The llama 3 herd of models.
\newblock \emph{arXiv preprint arXiv:2407.21783}, 2024.

\bibitem[Everitt et~al.(2011)Everitt, Landau, Leese, and Stahl]{everitt2011cluster}
Brian~S Everitt, Sabine Landau, Morven Leese, and Daniel Stahl.
\newblock Cluster analysis.
\newblock 2011.

\bibitem[Gao et~al.(2024)Gao, Li, Pan, Kuang, Ma, Qian, Wei, Zhang, Xie, Chen, et~al.]{gao2024agentscope}
Dawei Gao, Zitao Li, Xuchen Pan, Weirui Kuang, Zhijian Ma, Bingchen Qian, Fei Wei, Wenhao Zhang, Yuexiang Xie, Daoyuan Chen, et~al.
\newblock Agentscope: A flexible yet robust multi-agent platform.
\newblock \emph{arXiv preprint arXiv:2402.14034}, 2024.

\bibitem[Guo et~al.(2024)Guo, Chen, Wang, Chang, Pei, Chawla, Wiest, and Zhang]{guo2024large}
Taicheng Guo, Xiuying Chen, Yaqi Wang, Ruidi Chang, Shichao Pei, Nitesh~V Chawla, Olaf Wiest, and Xiangliang Zhang.
\newblock Large language model based multi-agents: A survey of progress and challenges.
\newblock \emph{arXiv preprint arXiv:2402.01680}, 2024.

\bibitem[Guu et~al.(2020)Guu, Lee, Tung, Pasupat, and Chang]{guu2020retrieval}
Kelvin Guu, Kenton Lee, Zora Tung, Panupong Pasupat, and Mingwei Chang.
\newblock Retrieval augmented language model pre-training.
\newblock In \emph{International conference on machine learning}, pp.\  3929--3938. PMLR, 2020.

\bibitem[Ho et~al.(2020)Ho, Nguyen, Sugawara, and Aizawa]{ho2020constructing}
Xanh Ho, Anh-Khoa~Duong Nguyen, Saku Sugawara, and Akiko Aizawa.
\newblock Constructing a multi-hop qa dataset for comprehensive evaluation of reasoning steps.
\newblock \emph{arXiv preprint arXiv:2011.01060}, 2020.

\bibitem[Hong et~al.(2023)Hong, Zheng, Chen, Cheng, Wang, Zhang, Wang, Yau, Lin, Zhou, et~al.]{hong2023metagpt}
Sirui Hong, Xiawu Zheng, Jonathan Chen, Yuheng Cheng, Jinlin Wang, Ceyao Zhang, Zili Wang, Steven Ka~Shing Yau, Zijuan Lin, Liyang Zhou, et~al.
\newblock Metagpt: Meta programming for multi-agent collaborative framework.
\newblock \emph{arXiv preprint arXiv:2308.00352}, 2023.

\bibitem[Iannelli et~al.(2024)Iannelli, Kuchipudi, and Dvorak]{iannelli2024sla}
Michael Iannelli, Sneha Kuchipudi, and Vera Dvorak.
\newblock Sla management in reconfigurable multi-agent rag: A systems approach to question answering.
\newblock \emph{arXiv preprint arXiv:2412.06832}, 2024.

\bibitem[Izacard \& Grave(2020)Izacard and Grave]{izacard2020leveraging}
Gautier Izacard and Edouard Grave.
\newblock Leveraging passage retrieval with generative models for open domain question answering.
\newblock \emph{arXiv preprint arXiv:2007.01282}, 2020.

\bibitem[Izacard et~al.(2021)Izacard, Caron, Hosseini, Riedel, Bojanowski, Joulin, and Grave]{izacard2021unsupervised}
Gautier Izacard, Mathilde Caron, Lucas Hosseini, Sebastian Riedel, Piotr Bojanowski, Armand Joulin, and Edouard Grave.
\newblock Unsupervised dense information retrieval with contrastive learning.
\newblock \emph{arXiv preprint arXiv:2112.09118}, 2021.

\bibitem[Jiang et~al.(2023)Jiang, Xu, Gao, Sun, Liu, Dwivedi-Yu, Yang, Callan, and Neubig]{jiang2023active}
Zhengbao Jiang, Frank~F Xu, Luyu Gao, Zhiqing Sun, Qian Liu, Jane Dwivedi-Yu, Yiming Yang, Jamie Callan, and Graham Neubig.
\newblock Active retrieval augmented generation.
\newblock \emph{arXiv preprint arXiv:2305.06983}, 2023.

\bibitem[Johnson et~al.(2021)Johnson, Gerlach, and Sáez-Trumper]{johnson2021classification}
Isaac Johnson, Martin Gerlach, and Diego Sáez-Trumper.
\newblock Language-agnostic topic classification for wikipedia.
\newblock \emph{WWW '21: Companion Proceedings of the Web Conference 2021}, pp.\  594–601, 2021.

\bibitem[Joshi et~al.(2024)Joshi, Sarwar, Varshney, Nag, Agrawal, and Naik]{joshi2024reaper}
Ashutosh Joshi, Sheikh~Muhammad Sarwar, Samarth Varshney, Sreyashi Nag, Shrivats Agrawal, and Juhi Naik.
\newblock Reaper: Reasoning based retrieval planning for complex rag systems.
\newblock In \emph{Proceedings of the 33rd ACM International Conference on Information and Knowledge Management}, pp.\  4621--4628, 2024.

\bibitem[Jung et~al.(2025)Jung, Jeong, and Huh]{jung2025federated}
Jincheol Jung, Hongju Jeong, and Eui-Nam Huh.
\newblock Federated learning and rag integration: a scalable approach for medical large language models.
\newblock In \emph{2025 International Conference on Artificial Intelligence in Information and Communication (ICAIIC)}, pp.\  0968--0973. IEEE, 2025.

\bibitem[Karpukhin et~al.(2020)Karpukhin, O{\u{g}}uz, Min, Lewis, Wu, Edunov, Chen, and Yih]{karpukhin2020dense}
Vladimir Karpukhin, Barlas O{\u{g}}uz, Sewon Min, Patrick Lewis, Ledell Wu, Sergey Edunov, Danqi Chen, and Wen-tau Yih.
\newblock Dense passage retrieval for open-domain question answering.
\newblock \emph{arXiv preprint arXiv:2004.04906}, 2020.

\bibitem[Kasai et~al.(2024)Kasai, Sakaguchi, Le~Bras, Asai, Yu, Radev, Smith, Choi, Inui, et~al.]{kasai2024realtime}
Jungo Kasai, Keisuke Sakaguchi, Ronan Le~Bras, Akari Asai, Xinyan Yu, Dragomir Radev, Noah~A Smith, Yejin Choi, Kentaro Inui, et~al.
\newblock Realtime qa: what's the answer right now?
\newblock \emph{Advances in Neural Information Processing Systems}, 36, 2024.

\bibitem[Kwiatkowski et~al.(2019)Kwiatkowski, Palomaki, Redfield, Collins, Parikh, Alberti, Epstein, Polosukhin, Devlin, Lee, et~al.]{kwiatkowski2019natural}
Tom Kwiatkowski, Jennimaria Palomaki, Olivia Redfield, Michael Collins, Ankur Parikh, Chris Alberti, Danielle Epstein, Illia Polosukhin, Jacob Devlin, Kenton Lee, et~al.
\newblock Natural questions: a benchmark for question answering research.
\newblock \emph{Transactions of the Association for Computational Linguistics}, 7:\penalty0 453--466, 2019.

\bibitem[Kwon et~al.(2023)Kwon, Li, Zhuang, Sheng, Zheng, Yu, Gonzalez, Zhang, and Stoica]{kwon2023efficient}
Woosuk Kwon, Zhuohan Li, Siyuan Zhuang, Ying Sheng, Lianmin Zheng, Cody~Hao Yu, Joseph~E. Gonzalez, Hao Zhang, and Ion Stoica.
\newblock Efficient memory management for large language model serving with pagedattention.
\newblock In \emph{Proceedings of the ACM SIGOPS 29th Symposium on Operating Systems Principles}, 2023.

\bibitem[Lewis et~al.(2020)Lewis, Perez, Piktus, Petroni, Karpukhin, Goyal, K{\"u}ttler, Lewis, Yih, Rockt{\"a}schel, et~al.]{lewis2020retrieval}
Patrick Lewis, Ethan Perez, Aleksandra Piktus, Fabio Petroni, Vladimir Karpukhin, Naman Goyal, Heinrich K{\"u}ttler, Mike Lewis, Wen-tau Yih, Tim Rockt{\"a}schel, et~al.
\newblock Retrieval-augmented generation for knowledge-intensive nlp tasks.
\newblock \emph{Advances in Neural Information Processing Systems}, 33:\penalty0 9459--9474, 2020.

\bibitem[Li et~al.(2024)Li, Xie, Li, Tsung, Ding, and Li]{li2024agent}
Ao~Li, Yuexiang Xie, Songze Li, Fugee Tsung, Bolin Ding, and Yaliang Li.
\newblock Agent-oriented planning in multi-agent systems.
\newblock \emph{arXiv preprint arXiv:2410.02189}, 2024.

\bibitem[Li et~al.(2023{\natexlab{a}})Li, Hammoud, Itani, Khizbullin, and Ghanem]{li2023camel}
Guohao Li, Hasan Hammoud, Hani Itani, Dmitrii Khizbullin, and Bernard Ghanem.
\newblock Camel: Communicative agents for" mind" exploration of large language model society.
\newblock \emph{Advances in Neural Information Processing Systems}, 36:\penalty0 51991--52008, 2023{\natexlab{a}}.

\bibitem[Li et~al.(2023{\natexlab{b}})Li, Zhu, Li, Yin, Sun, and Qiu]{li2023llatrieval}
Xiaonan Li, Changtai Zhu, Linyang Li, Zhangyue Yin, Tianxiang Sun, and Xipeng Qiu.
\newblock Llatrieval: Llm-verified retrieval for verifiable generation.
\newblock \emph{arXiv preprint arXiv:2311.07838}, 2023{\natexlab{b}}.

\bibitem[Li et~al.(2025)Li, Zhang, Yang, Huang, Wu, Luo, Bei, Zou, Luo, Zhao, et~al.]{li2025towards}
Yangning Li, Weizhi Zhang, Yuyao Yang, Wei-Chieh Huang, Yaozu Wu, Junyu Luo, Yuanchen Bei, Henry~Peng Zou, Xiao Luo, Yusheng Zhao, et~al.
\newblock Towards agentic rag with deep reasoning: A survey of rag-reasoning systems in llms.
\newblock \emph{arXiv preprint arXiv:2507.09477}, 2, 2025.

\bibitem[Liu(2022)]{LiuLlamaIndex2022}
Jerry Liu.
\newblock {LlamaIndex}, 11 2022.
\newblock URL \url{https://github.com/jerryjliu/llama_index}.

\bibitem[Liu et~al.(2026)Liu, Zhang, Yu, Xiong, He, Wu, Jung, Fredrikson, Wang, and Gao]{liu2026vision}
Xiaoze Liu, Ruowang Zhang, Weichen Yu, Siheng Xiong, Liu He, Feijie Wu, Hoin Jung, Matt Fredrikson, Xiaoqian Wang, and Jing Gao.
\newblock The vision wormhole: Latent-space communication in heterogeneous multi-agent systems.
\newblock \emph{arXiv preprint arXiv:2602.15382}, 2026.

\bibitem[Lu et~al.(2023)Lu, Yuan, Lin, Lin, Yuan, Zhou, and Zhou]{lu2023routing}
Keming Lu, Hongyi Yuan, Runji Lin, Junyang Lin, Zheng Yuan, Chang Zhou, and Jingren Zhou.
\newblock Routing to the expert: Efficient reward-guided ensemble of large language models.
\newblock \emph{arXiv preprint arXiv:2311.08692}, 2023.

\bibitem[Lu et~al.(2024{\natexlab{a}})Lu, Peng, Cheng, Galley, Chang, Wu, Zhu, and Gao]{lu2024chameleon}
Pan Lu, Baolin Peng, Hao Cheng, Michel Galley, Kai-Wei Chang, Ying~Nian Wu, Song-Chun Zhu, and Jianfeng Gao.
\newblock Chameleon: Plug-and-play compositional reasoning with large language models.
\newblock \emph{Advances in Neural Information Processing Systems}, 36, 2024{\natexlab{a}}.

\bibitem[Lu et~al.(2024{\natexlab{b}})Lu, Yu, and Khashabi]{lu2024gear}
Yining Lu, Haoping Yu, and Daniel Khashabi.
\newblock Gear: Augmenting language models with generalizable and efficient tool resolution.
\newblock In \emph{Proceedings of the 18th Conference of the European Chapter of the Association for Computational Linguistics (Volume 1: Long Papers)}, pp.\  112--138, 2024{\natexlab{b}}.

\bibitem[Ma et~al.(2023)Ma, Gong, He, Zhao, and Duan]{ma2023query}
Xinbei Ma, Yeyun Gong, Pengcheng He, Hai Zhao, and Nan Duan.
\newblock Query rewriting for retrieval-augmented large language models.
\newblock \emph{arXiv preprint arXiv:2305.14283}, 2023.

\bibitem[Mallen et~al.(2022)Mallen, Asai, Zhong, Das, Khashabi, and Hajishirzi]{mallen2022not}
Alex Mallen, Akari Asai, Victor Zhong, Rajarshi Das, Daniel Khashabi, and Hannaneh Hajishirzi.
\newblock When not to trust language models: Investigating effectiveness of parametric and non-parametric memories.
\newblock \emph{arXiv preprint arXiv:2212.10511}, 2022.

\bibitem[Mei et~al.(2025)Mei, Xu, Guo, Lin, and Zhang]{mei2025omnirouter}
Kai Mei, Wujiang Xu, Minghao Guo, Shuhang Lin, and Yongfeng Zhang.
\newblock Omnirouter: Budget and performance controllable multi-llm routing.
\newblock \emph{ACM SIGKDD Explorations Newsletter}, 27\penalty0 (2):\penalty0 107--116, 2025.

\bibitem[Murtagh \& Contreras(2012)Murtagh and Contreras]{murtagh2012algorithms}
Fionn Murtagh and Pedro Contreras.
\newblock Algorithms for hierarchical clustering: an overview.
\newblock \emph{Wiley interdisciplinary reviews: data mining and knowledge discovery}, 2\penalty0 (1):\penalty0 86--97, 2012.

\bibitem[Ong et~al.(2024)Ong, Almahairi, Wu, Chiang, Wu, Gonzalez, Kadous, and Stoica]{ong2024routellm}
Isaac Ong, Amjad Almahairi, Vincent Wu, Wei-Lin Chiang, Tianhao Wu, Joseph~E Gonzalez, M~Waleed Kadous, and Ion Stoica.
\newblock Routellm: Learning to route llms with preference data.
\newblock \emph{arXiv preprint arXiv:2406.18665}, 2024.

\bibitem[Patil et~al.(2024)Patil, Zhang, Wang, and Gonzalez]{patil2024gorilla}
Shishir~G Patil, Tianjun Zhang, Xin Wang, and Joseph~E Gonzalez.
\newblock Gorilla: Large language model connected with massive apis.
\newblock \emph{Advances in Neural Information Processing Systems}, 37:\penalty0 126544--126565, 2024.

\bibitem[Press et~al.(2022)Press, Zhang, Min, Schmidt, Smith, and Lewis]{press2022measuring}
Ofir Press, Muru Zhang, Sewon Min, Ludwig Schmidt, Noah~A Smith, and Mike Lewis.
\newblock Measuring and narrowing the compositionality gap in language models.
\newblock \emph{arXiv preprint arXiv:2210.03350}, 2022.

\bibitem[Robertson et~al.(2004)Robertson, Zaragoza, and Taylor]{robertson2004simple}
Stephen Robertson, Hugo Zaragoza, and Michael Taylor.
\newblock Simple bm25 extension to multiple weighted fields.
\newblock In \emph{Proceedings of the thirteenth ACM international conference on Information and knowledge management}, pp.\  42--49, 2004.

\bibitem[Roucher(2024)]{huggingfaceRAGAgent}
Aymeric Roucher.
\newblock {Agentic RAG}: turbocharge your {RAG} with query reformulation and self-query.
\newblock \url{https://huggingface.co/learn/cookbook/en/agent_rag}, 2024.
\newblock URL \url{https://huggingface.co/learn/cookbook/en/agent_rag}.
\newblock Accessed: 2025-02-15.

\bibitem[Shao et~al.(2023)Shao, Gong, Shen, Huang, Duan, and Chen]{shao2023enhancing}
Zhihong Shao, Yeyun Gong, Yelong Shen, Minlie Huang, Nan Duan, and Weizhu Chen.
\newblock Enhancing retrieval-augmented large language models with iterative retrieval-generation synergy.
\newblock \emph{arXiv preprint arXiv:2305.15294}, 2023.

\bibitem[Sharma et~al.(2019)Sharma, Batra, et~al.]{sharma2019comparative}
Shweta Sharma, Neha Batra, et~al.
\newblock Comparative study of single linkage, complete linkage, and ward method of agglomerative clustering.
\newblock In \emph{2019 international conference on machine learning, big data, cloud and parallel computing (COMITCon)}, pp.\  568--573. IEEE, 2019.

\bibitem[Shi et~al.(2023)Shi, Min, Yasunaga, Seo, James, Lewis, Zettlemoyer, and Yih]{shi2023replug}
Weijia Shi, Sewon Min, Michihiro Yasunaga, Minjoon Seo, Rich James, Mike Lewis, Luke Zettlemoyer, and Wen-tau Yih.
\newblock Replug: Retrieval-augmented black-box language models.
\newblock \emph{arXiv preprint arXiv:2301.12652}, 2023.

\bibitem[Shi et~al.(2025)Shi, Wan, Huang, Zhang, Shao, Ye, and Yang]{shi2025privacy}
Zitong Shi, Guancheng Wan, Wenke Huang, Guibin Zhang, Jiawei Shao, Mang Ye, and Carl Yang.
\newblock Privacy-enhancing paradigms within federated multi-agent systems.
\newblock \emph{arXiv preprint arXiv:2503.08175}, 2025.

\bibitem[Singh et~al.(2025)Singh, Ehtesham, Kumar, and Khoei]{singh2025agentic}
Aditi Singh, Abul Ehtesham, Saket Kumar, and Tala~Talaei Khoei.
\newblock Agentic retrieval-augmented generation: A survey on agentic rag.
\newblock \emph{arXiv preprint arXiv:2501.09136}, 2025.

\bibitem[Su et~al.(2024)Su, Tang, Ai, Wu, and Liu]{su2024dragin}
Weihang Su, Yichen Tang, Qingyao Ai, Zhijing Wu, and Yiqun Liu.
\newblock Dragin: Dynamic retrieval augmented generation based on the real-time information needs of large language models.
\newblock \emph{arXiv preprint arXiv:2403.10081}, 2024.

\bibitem[Tang \& Yang(2024)Tang and Yang]{tang2024multihop}
Yixuan Tang and Yi~Yang.
\newblock Multihop-rag: Benchmarking retrieval-augmented generation for multi-hop queries.
\newblock \emph{arXiv preprint arXiv:2401.15391}, 2024.

\bibitem[Touvron et~al.(2023{\natexlab{a}})Touvron, Lavril, Izacard, Martinet, Lachaux, Lacroix, Rozi{\`e}re, Goyal, Hambro, Azhar, et~al.]{touvron2023llama}
Hugo Touvron, Thibaut Lavril, Gautier Izacard, Xavier Martinet, Marie-Anne Lachaux, Timoth{\'e}e Lacroix, Baptiste Rozi{\`e}re, Naman Goyal, Eric Hambro, Faisal Azhar, et~al.
\newblock Llama: Open and efficient foundation language models.
\newblock \emph{arXiv preprint arXiv:2302.13971}, 2023{\natexlab{a}}.

\bibitem[Touvron et~al.(2023{\natexlab{b}})Touvron, Martin, Stone, Albert, Almahairi, Babaei, Bashlykov, Batra, Bhargava, Bhosale, et~al.]{touvron2023llama2}
Hugo Touvron, Louis Martin, Kevin Stone, Peter Albert, Amjad Almahairi, Yasmine Babaei, Nikolay Bashlykov, Soumya Batra, Prajjwal Bhargava, Shruti Bhosale, et~al.
\newblock Llama 2: Open foundation and fine-tuned chat models.
\newblock \emph{arXiv preprint arXiv:2307.09288}, 2023{\natexlab{b}}.

\bibitem[Trivedi et~al.(2022{\natexlab{a}})Trivedi, Balasubramanian, Khot, and Sabharwal]{trivedi2022interleaving}
Harsh Trivedi, Niranjan Balasubramanian, Tushar Khot, and Ashish Sabharwal.
\newblock Interleaving retrieval with chain-of-thought reasoning for knowledge-intensive multi-step questions.
\newblock \emph{arXiv preprint arXiv:2212.10509}, 2022{\natexlab{a}}.

\bibitem[Trivedi et~al.(2022{\natexlab{b}})Trivedi, Balasubramanian, Khot, and Sabharwal]{trivedi2022musique}
Harsh Trivedi, Niranjan Balasubramanian, Tushar Khot, and Ashish Sabharwal.
\newblock Musique: Multihop questions via single-hop question composition.
\newblock \emph{Transactions of the Association for Computational Linguistics}, 10:\penalty0 539--554, 2022{\natexlab{b}}.

\bibitem[Verma et~al.(2024)Verma, Midigeshi, Sinha, Solin, Natarajan, and Sharma]{verma2024plan}
Prakhar Verma, Sukruta~Prakash Midigeshi, Gaurav Sinha, Arno Solin, Nagarajan Natarajan, and Amit Sharma.
\newblock Plan $\times$ rag: Planning-guided retrieval augmented generation.
\newblock \emph{arXiv preprint arXiv:2410.20753}, 2024.

\bibitem[Wang et~al.(2026)Wang, Li, Zhang, Chen, Chen, Jian, Zhang, and Hu]{wang2026icl}
Chenxu Wang, Hao Li, Yiqun Zhang, Linyao Chen, Jianhao Chen, Ping Jian, Qiaosheng Zhang, and Shuyue Hu.
\newblock Icl-router: In-context learned model representations for llm routing.
\newblock In \emph{Proceedings of the AAAI Conference on Artificial Intelligence}, volume~40, pp.\  33413--33421, 2026.

\bibitem[Wei et~al.(2025)Wei, Yang, Chen, Zhao, Dernoncourt, Rossi, and Eldardiry]{wei2025learning}
Wang Wei, Tiankai Yang, Hongjie Chen, Yue Zhao, Franck Dernoncourt, Ryan~A Rossi, and Hoda Eldardiry.
\newblock Learning to route llms from bandit feedback: One policy, many trade-offs.
\newblock \emph{arXiv preprint arXiv:2510.07429}, 2025.

\bibitem[Weng(2023)]{agent_intro}
Lilian Weng.
\newblock {LLM} powered autonomous agents.
\newblock \url{https://lilianweng.github.io/posts/2023-06-23-agent/}, 2023.
\newblock URL \url{https://lilianweng.github.io/posts/2023-06-23-agent/}.
\newblock Accessed: 2025-01-09.

\bibitem[Wu et~al.(2024{\natexlab{a}})Wu, Li, Li, Ding, and Gao]{wu2024fedbiot}
Feijie Wu, Zitao Li, Yaliang Li, Bolin Ding, and Jing Gao.
\newblock Fedbiot: Llm local fine-tuning in federated learning without full model.
\newblock In \emph{Proceedings of the 30th ACM SIGKDD conference on knowledge discovery and data mining}, pp.\  3345--3355, 2024{\natexlab{a}}.

\bibitem[Wu et~al.(2024{\natexlab{b}})Wu, Liu, Wang, Wang, Su, and Gao]{wu2024towards}
Feijie Wu, Xiaoze Liu, Haoyu Wang, Xingchen Wang, Lu~Su, and Jing Gao.
\newblock Towards federated rlhf with aggregated client preference for llms.
\newblock \emph{arXiv preprint arXiv:2407.03038}, 2024{\natexlab{b}}.

\bibitem[Wu et~al.(2025)Wu, Zhu, Zhang, Chatterjee, Zhu, Mo, Luo, and Gao]{wu2025portool}
Feijie Wu, Weiwu Zhu, Yuxiang Zhang, Soumya Chatterjee, Jiarong Zhu, Fan Mo, Rodin Luo, and Jing Gao.
\newblock Portool: Tool-use llm training with rewarded tree.
\newblock \emph{arXiv preprint arXiv:2510.26020}, 2025.

\bibitem[Wu et~al.(2023)Wu, Bansal, Zhang, Wu, Zhang, Zhu, Li, Jiang, Zhang, and Wang]{wu2023autogen}
Qingyun Wu, Gagan Bansal, Jieyu Zhang, Yiran Wu, Shaokun Zhang, Erkang Zhu, Beibin Li, Li~Jiang, Xiaoyun Zhang, and Chi Wang.
\newblock Autogen: Enabling next-gen llm applications via multi-agent conversation framework.
\newblock \emph{arXiv preprint arXiv:2308.08155}, 2023.

\bibitem[Wu et~al.(2024{\natexlab{c}})Wu, Xiong, Cui, Wu, Chen, Yuan, Huang, Liu, Kuo, Guan, et~al.]{wu2024retrieval}
Shangyu Wu, Ying Xiong, Yufei Cui, Haolun Wu, Can Chen, Ye~Yuan, Lianming Huang, Xue Liu, Tei-Wei Kuo, Nan Guan, et~al.
\newblock Retrieval-augmented generation for natural language processing: A survey.
\newblock \emph{arXiv preprint arXiv:2407.13193}, 2024{\natexlab{c}}.

\bibitem[Xiong et~al.(2024)Xiong, Jin, Lu, and Zhang]{xiong2024benchmarking}
Guangzhi Xiong, Qiao Jin, Zhiyong Lu, and Aidong Zhang.
\newblock Benchmarking retrieval-augmented generation for medicine.
\newblock \emph{arXiv preprint arXiv:2402.13178}, 2024.

\bibitem[Xu et~al.(2023)Xu, Ping, Wu, McAfee, Zhu, Liu, Subramanian, Bakhturina, Shoeybi, and Catanzaro]{xu2023retrieval}
Peng Xu, Wei Ping, Xianchao Wu, Lawrence McAfee, Chen Zhu, Zihan Liu, Sandeep Subramanian, Evelina Bakhturina, Mohammad Shoeybi, and Bryan Catanzaro.
\newblock Retrieval meets long context large language models.
\newblock \emph{arXiv preprint arXiv:2310.03025}, 2023.

\bibitem[Yan et~al.(2024)Yan, Gu, Zhu, and Ling]{yan2024corrective}
Shi-Qi Yan, Jia-Chen Gu, Yun Zhu, and Zhen-Hua Ling.
\newblock Corrective retrieval augmented generation.
\newblock \emph{arXiv preprint arXiv:2401.15884}, 2024.

\bibitem[Yang et~al.(2024{\natexlab{a}})Yang, Yang, Zhang, Hui, Zheng, Yu, Li, Liu, Huang, Wei, et~al.]{yang2024qwen2}
An~Yang, Baosong Yang, Beichen Zhang, Binyuan Hui, Bo~Zheng, Bowen Yu, Chengyuan Li, Dayiheng Liu, Fei Huang, Haoran Wei, et~al.
\newblock Qwen2. 5 technical report.
\newblock \emph{arXiv preprint arXiv:2412.15115}, 2024{\natexlab{a}}.

\bibitem[Yang et~al.(2024{\natexlab{b}})Yang, Rao, Chen, Guo, Zhang, Yang, and Zhang]{yang2024rag}
Diji Yang, Jinmeng Rao, Kezhen Chen, Xiaoyuan Guo, Yawen Zhang, Jie Yang, and Yi~Zhang.
\newblock Im-rag: Multi-round retrieval-augmented generation through learning inner monologues.
\newblock In \emph{Proceedings of the 47th International ACM SIGIR Conference on Research and Development in Information Retrieval}, pp.\  730--740, 2024{\natexlab{b}}.

\bibitem[Yang et~al.(2018)Yang, Qi, Zhang, Bengio, Cohen, Salakhutdinov, and Manning]{yang2018hotpotqa}
Zhilin Yang, Peng Qi, Saizheng Zhang, Yoshua Bengio, William~W Cohen, Ruslan Salakhutdinov, and Christopher~D Manning.
\newblock Hotpotqa: A dataset for diverse, explainable multi-hop question answering.
\newblock \emph{arXiv preprint arXiv:1809.09600}, 2018.

\bibitem[Ye et~al.(2025)Ye, Tang, Ge, Du, Yin, Chen, and Shao]{ye2025mas}
Rui Ye, Shuo Tang, Rui Ge, Yaxin Du, Zhenfei Yin, Siheng Chen, and Jing Shao.
\newblock Mas-gpt: Training llms to build llm-based multi-agent systems.
\newblock \emph{arXiv preprint arXiv:2503.03686}, 2025.

\bibitem[Yu(2022)]{yu2022retrieval}
Wenhao Yu.
\newblock Retrieval-augmented generation across heterogeneous knowledge.
\newblock In \emph{Proceedings of the 2022 conference of the North American chapter of the association for computational linguistics: human language technologies: student research workshop}, pp.\  52--58, 2022.

\bibitem[Yu et~al.(2024)Yu, Ping, Liu, Wang, You, Zhang, Shoeybi, and Catanzaro]{yu2024rankrag}
Yue Yu, Wei Ping, Zihan Liu, Boxin Wang, Jiaxuan You, Chao Zhang, Mohammad Shoeybi, and Bryan Catanzaro.
\newblock Rankrag: Unifying context ranking with retrieval-augmented generation in llms.
\newblock \emph{arXiv preprint arXiv:2407.02485}, 2024.

\bibitem[Yuan et~al.(2025)Yuan, Song, Chen, Tan, Shen, Ren, Li, and Yang]{yuan2025easytool}
Siyu Yuan, Kaitao Song, Jiangjie Chen, Xu~Tan, Yongliang Shen, Kan Ren, Dongsheng Li, and Deqing Yang.
\newblock Easytool: Enhancing llm-based agents with concise tool instruction.
\newblock In \emph{Proceedings of the 2025 Conference of the Nations of the Americas Chapter of the Association for Computational Linguistics: Human Language Technologies (Volume 1: Long Papers)}, pp.\  951--972, 2025.

\bibitem[Yue et~al.(2025)Yue, Zhang, Liu, Wan, Wang, Cheng, and Qi]{yue2025masrouter}
Yanwei Yue, Guibin Zhang, Boyang Liu, Guancheng Wan, Kun Wang, Dawei Cheng, and Yiyan Qi.
\newblock Masrouter: Learning to route llms for multi-agent systems.
\newblock \emph{arXiv preprint arXiv:2502.11133}, 2025.

\bibitem[Zhang et~al.(2025{\natexlab{a}})Zhang, Li, Chen, Wang, and He]{zhang2025belle}
Taolin Zhang, Dongyang Li, Qizhou Chen, Chengyu Wang, and Xiaofeng He.
\newblock Belle: A bi-level multi-agent reasoning framework for multi-hop question answering.
\newblock \emph{arXiv preprint arXiv:2505.11811}, 2025{\natexlab{a}}.

\bibitem[Zhang et~al.(2024)Zhang, Zhang, Long, Xie, Dai, Tang, Lin, Yang, Xie, Huang, et~al.]{zhang2024mgte}
Xin Zhang, Yanzhao Zhang, Dingkun Long, Wen Xie, Ziqi Dai, Jialong Tang, Huan Lin, Baosong Yang, Pengjun Xie, Fei Huang, et~al.
\newblock mgte: Generalized long-context text representation and reranking models for multilingual text retrieval.
\newblock \emph{arXiv preprint arXiv:2407.19669}, 2024.

\bibitem[Zhang et~al.(2025{\natexlab{b}})Zhang, Shi, Yuan, Wang, Ma, Murugesan, Galassi, Zhang, and Ye]{zhang2025agentrouter}
Zheyuan Zhang, Kaiwen Shi, Zhengqing Yuan, Zehong Wang, Tianyi Ma, Keerthiram Murugesan, Vincent Galassi, Chuxu Zhang, and Yanfang Ye.
\newblock Agentrouter: A knowledge-graph-guided llm router for collaborative multi-agent question answering.
\newblock \emph{arXiv preprint arXiv:2510.05445}, 2025{\natexlab{b}}.

\bibitem[Zhao(2024)]{zhao2024frag}
Dongfang Zhao.
\newblock Frag: Toward federated vector database management for collaborative and secure retrieval-augmented generation.
\newblock \emph{arXiv preprint arXiv:2410.13272}, 2024.

\bibitem[Zhao et~al.(2023)Zhao, Wang, Zhang, Jin, Zhu, Chen, and Xie]{zhao2023competeai}
Qinlin Zhao, Jindong Wang, Yixuan Zhang, Yiqiao Jin, Kaijie Zhu, Hao Chen, and Xing Xie.
\newblock Competeai: Understanding the competition behaviors in large language model-based agents.
\newblock \emph{arXiv preprint arXiv:2310.17512}, 2023.

\bibitem[Zheng et~al.(2026)Zheng, Zhang, Ma, Yu, Zhu, Dong, and Zhu]{zheng2026skillrouter}
YanZhao Zheng, ZhenTao Zhang, Chao Ma, YuanQiang Yu, JiHuan Zhu, Baohua Dong, and Hangcheng Zhu.
\newblock Skillrouter: Retrieve-and-rerank skill selection for llm agents at scale.
\newblock \emph{arXiv preprint arXiv:2603.22455}, 2026.

\bibitem[Zhuang et~al.(2024)Zhuang, Zhang, Cheng, Yang, Liu, Huang, Lin, Rajmohan, Zhang, and Zhang]{zhuang2024efficientrag}
Ziyuan Zhuang, Zhiyang Zhang, Sitao Cheng, Fangkai Yang, Jia Liu, Shujian Huang, Qingwei Lin, Saravan Rajmohan, Dongmei Zhang, and Qi~Zhang.
\newblock Efficientrag: Efficient retriever for multi-hop question answering.
\newblock \emph{arXiv preprint arXiv:2408.04259}, 2024.

\end{thebibliography}
\bibliographystyle{colm2026_conference}

\newpage

\appendix

\section{More Details of \algo} \label{apdx:rirs_more}

This section complements \cref{sec: routing} by presenting implementation-level details and additional analysis that are omitted from the main text due to space constraints. We focus on (i) why centroid-based knowledge clustering is a reliable routing primitive, and (ii) how the central server operationalizes \algo through lightweight, knowledge-free modules (evaluation, summarization, and query refinement) that interact with routed RAG agents using quoted evidence. Throughout, we follow the notation and workflow in the main text.

\subsection{Optimization of \cref{eq:cluster-maximin}: Approximate Max-Min Similarity Clustering}
\label{apdx:opt_algo}

\begin{algorithm}[p]
\caption{Approximate Max-Min Similarity Clustering}
\label{algo}
\begin{algorithmic}[1]
\Require Embeddings $E=\{e_1,\dots,e_m\}$, target number of clusters $n$, refinement rounds $R\ge 0$
\Ensure Partition $\mathcal{C}=\{C_1,\dots,C_n\}$ and centroids $\{\bar{c}_1,\dots,\bar{c}_n\}$

\State Normalize each embedding: $\hat{e}_i \gets e_i / \|e_i\|$ for all $i\in[m]$
\State Compute cosine similarity matrix $S\in\mathbb{R}^{m\times m}$, where $S_{ij}=\hat{e}_i^\top \hat{e}_j$
\State Initialize $\mathcal{C}\gets \big\{\{1\},\{2\},\dots,\{m\}\big\}$
\State Define cluster bottleneck
\[
b(C)\gets
\begin{cases}
+\infty, & |C|\le 1,\\[1mm]
\min_{i,j\in C,\ i<j} S_{ij}, & |C|\ge 2
\end{cases}
\]

\Statex
\State \textbf{Stage I: Cached complete-link clustering}
\State Initialize $\mathrm{active}(A)\gets \textbf{true}$ for all $A\in\mathcal{C}$
\State Initialize $G[\{i\},\{j\}] \gets S_{ij}$ for all $i\neq j$
\State Initialize an empty max-heap $H$
\ForAll{unordered pairs of distinct clusters $(A,B)$ with $A,B\in\mathcal{C}$}
    \State Push $(G[A,B], A, B)$ into $H$
\EndFor

\While{number of active clusters $> n$}
    \Repeat
        \State Pop the top entry $(\gamma, A, B)$ from $H$
    \Until{$\mathrm{active}(A)$ and $\mathrm{active}(B)$}
    \State $M \gets A \cup B$
    \State $\mathrm{active}(A)\gets \textbf{false},\ \mathrm{active}(B)\gets \textbf{false},\ \mathrm{active}(M)\gets \textbf{true}$
    \ForAll{active clusters $C$ with $C \neq M$}
        \State $G[M,C] \gets \min\{G[A,C],\,G[B,C]\}$; set $G[C,M]\gets G[M,C]$
        \State Push $(G[M,C], M, C)$ into $H$
    \EndFor
\EndWhile
\State $\mathcal{C} \gets \{A\in\mathcal{C} : \mathrm{active}(A)\}$

\Statex
\State \textbf{Stage II: Worst-pair local refinements}
\For{$r=1$ to $R$}
    \State $A \gets \arg\min_{C\in\mathcal{C}} b(C)$ and $(a,b)\gets \arg\min_{i,j\in A,\ i<j} S_{ij}$
    \State $\mathrm{bestScore}\gets b(A)$, \ $\mathrm{bestMove}\gets \varnothing$
    \ForAll{$p\in\{a,b\}$ and $D\in\mathcal{C}\setminus\{A\}$}
        \State $\mathrm{newScore}\gets \min\!\Big\{ b(A\setminus\{p\}),\ \min\big(b(D), \min_{j\in D} S_{pj}\big),\ \min_{X\in\mathcal{C}\setminus\{A,D\}} b(X)\Big\}$
        \If{$\mathrm{newScore}>\mathrm{bestScore}$}
            \State $\mathrm{bestScore}\gets \mathrm{newScore}$, \ $\mathrm{bestMove}\gets (p,A,D)$
        \EndIf
    \EndFor
    \If{$\mathrm{bestMove}=\varnothing$}
        \State \textbf{break}
    \EndIf
    \State Apply $\mathrm{bestMove}$ by moving $p$ from $A$ to $D$
\EndFor

\State Relabel the clusters in $\mathcal{C}$ as $\{C_1,\dots,C_n\}$

\Statex
\State \textbf{Stage III: Centroid computation}
\For{$k=1$ to $n$}
    \State Compute centroid:
    \[
    \bar{c}_k \gets \frac{1}{|C_k|}\sum_{i\in C_k} e_i
    \]
\EndFor

\State \Return $\mathcal{C}=\{C_1,\dots,C_n\}$ and $\{\bar{c}_1,\dots,\bar{c}_n\}$
\end{algorithmic}
\end{algorithm}

\cref{eq:cluster-maximin} seeks a partition of the local embeddings into $n$ non-empty clusters that maximizes the smallest within-cluster cosine similarity. Exact optimization over all partitions is combinatorial. We therefore adopt a two-stage greedy approximation, summarized in \cref{algo}. The first stage is motivated by classic work on hierarchical clustering and greedy complete-linkage agglomeration \cite{everitt2011cluster,murtagh2012algorithms,sharma2019comparative}. The second stage adds an $R$-round local refinement to further improve the max-min objective.

Let $\hat e_i=e_i/\|e_i\|$ denote the normalized embedding of knowledge piece $i$, and let $S_{ij}=\hat e_i^\top \hat e_j$ be the cosine similarity between normalized embeddings. For any non-empty cluster $C$, define its bottleneck similarity by
\[
b(C):=
\begin{cases}
+\infty, & |C|\le 1,\\[1mm]
\min_{i,j\in C,\ i<j} S_{ij}, & |C|\ge 2.
\end{cases}
\]
For a partition $\mathcal C=\{C_1,\dots,C_n\}$, define the partition bottleneck objective
\[
B(\mathcal C):=\min_{k\in[n]} b(C_k).
\]
Then \cref{eq:cluster-maximin} is equivalent to maximizing $B(\mathcal C)$ over all partitions into $n$ non-empty clusters.

\paragraph{Stage I: Cached complete-link clustering.}
We initialize each embedding as a singleton cluster. For two current clusters $A$ and $B$, define their merge score by
\[
g(A,B):=\min_{i\in A,\ j\in B} S_{ij}.
\]
At each step, Stage I merges the pair of active clusters with the largest value of $g(A,B)$. This rule is the complete-link analogue of greedy agglomeration: it favors merges whose worst cross-cluster pair is as strong as possible, and thus delays deterioration of the smallest within-cluster similarity. To make this stage efficient, we cache inter-cluster bottleneck scores and maintain candidate merges in a max-heap. After a merge $M=A\cup B$, the score between $M$ and any other active cluster $C$ is updated by
\[
g(M,C)=\min\{g(A,C),\,g(B,C)\},
\]
which is the standard complete-link recurrence. Stage I stops once only $n$ active clusters remain.

The first stage is greedy and does not guarantee a globally optimal solution to \cref{eq:cluster-maximin}. Its role is to construct a strong initialization that already respects the max-min structure of the objective.

\paragraph{Stage II: Worst-pair local refinement.}
The agglomerative initialization can still be suboptimal because early merge decisions are irreversible. We therefore perform up to $R$ rounds of local refinement. At each round, we identify the current worst cluster
\[
A^\star \in \arg\min_{C\in\mathcal C} b(C),
\]
and then identify its worst pair
\[
(a^\star,b^\star)\in \arg\min_{i,j\in A^\star,\ i<j} S_{ij}.
\]
We consider moving one endpoint of this worst pair, namely $a^\star$ or $b^\star$, from $A^\star$ to another cluster $D\in \mathcal C\setminus\{A^\star\}$. Among all such candidate moves, we select the one that yields the largest improvement in $B(\mathcal C)$, and we accept it only if the objective increases strictly. If no improving move exists, the refinement stage terminates early.

This refinement is motivated by the observation that the max-min objective is controlled by the weakest cluster and, within that cluster, by its weakest pair. Focusing local search on this bottleneck is therefore a direct way to improve the overall partition quality. The procedure is easiest to describe under a mild generic-position assumption that distinct embedding pairs have distinct cosine similarities. Under this assumption, the worst cluster and the worst pair at each round are uniquely defined. In practice, this assumption is natural for dense embeddings of different knowledge pieces, since exact equality between two pairwise cosine similarities is rare. If ties do occur, they can be broken arbitrarily without changing the procedure.

\paragraph{Centroid computation.}
After the final partition $\mathcal C=\{C_1,\dots,C_n\}$ is obtained, the agent computes one centroid for each cluster:
\[
\bar c_k = \frac{1}{|C_k|}\sum_{i\in C_k} e_i,\qquad k=1,\dots,n.
\]
Only these centroids are uploaded to the server for routing.

\paragraph{Complexity analysis.}
The space complexity is $O(m^2)$, dominated by the pairwise similarity matrix and the cached inter-cluster bottleneck scores.

The time complexity is $O(m^2(R+\log m))$. Computing the pairwise cosine similarity matrix requires $O(m^2)$ time. In Stage I, the heap initialization and the sequence of greedy complete-link merges together cost $O(m^2\log m)$. In Stage II, each refinement round evaluates candidate moves for the current worst pair against the remaining clusters, which leads to $O(m^2)$ time per round in the worst case. Over $R$ rounds, this gives $O(Rm^2)$. Combining both stages yields the total complexity
\[
O\big(m^2(R+\log m)\big).
\]

In our implementation, we use this two-stage approximation because it is simple, deterministic given tie-breaking, and directly targets the bottleneck structure of \cref{eq:cluster-maximin} without introducing any additional learned parameters.

\vspace{-5px}
\subsection{Effectiveness of Knowledge Summarization} \label{apdx:clustering}

This subsection provides additional justification for centroid-based routing in \algo. In the offline stage (\cref{subsec:clustering}), each RAG agent partitions its local knowledge-piece embeddings into disjoint clusters and uploads only the corresponding centroids to the server. At inference time, the server routes a user query by retrieving the most similar centroids and selecting the associated agents/clusters. The key intuition is that if a cluster is semantically coherent, then its centroid remains a reliable proxy for locating that cluster (and thus the responsible RAG agent) using cosine similarity.

We formalize this intuition in a similarity-bounded (threshold-based) clustering setting and show that centroid-based search is \emph{bounded}: a query cannot be simultaneously very close to a relevant knowledge piece and arbitrarily far from the centroid of the piece's cluster. This bound clarifies when centroid routing is reliable and how clustering quality (coherence) impacts routing robustness.

\vspace{-5px}
\paragraph{Setup and Notation.}
Consider an RAG agent with $m$ knowledge pieces whose embeddings are $\{e_1,\dots,e_m\}$. Let $\{c_1,\dots,c_n\}$ be a partition of these embeddings into $n$ non-empty clusters. For a cluster $c_i$, its centroid is
$\bar{c}_i = \frac{1}{|c_i|}\sum_{\hat{e}\in c_i}\hat{e}$.
Let $\mathrm{sim}(\cdot,\cdot)$ denote cosine similarity. We say $c_i$ is \emph{$s$-coherent} if for any two distinct embeddings $e_a,e_b\in c_i$, $\mathrm{sim}(e_a,e_b)\ge s$ with $s>0$.

\begin{theorem}[Embedding-to-centroid bound] \label{theo:embed_cen_sim_apdx}
If a cluster $c_i$ is $s$-coherent, then for any embedding $e\in c_i$ and its centroid $\bar{c}_i$, we have $\mathrm{sim}(e,\bar{c}_i)\ge s$.
\end{theorem}
\begin{proof}
By definition, $\bar{c}_i=\frac{1}{|c_i|}\sum_{\hat{e}\in c_i}\hat{e}$. For any $e\in c_i$,
\begin{align*}
\mathrm{sim}(e,\bar{c}_i)
=\frac{e\cdot \bar{c}_i}{\|e\|\cdot\|\bar{c}_i\|}
=\frac{1}{|c_i|}\sum_{\hat{e}\in c_i}\frac{e\cdot \hat{e}}{\|e\|\cdot\|\bar{c}_i\|} =\frac{1}{|c_i|}\sum_{\hat{e}\in c_i}\frac{\|\hat{e}\|\cdot \mathrm{sim}(e,\hat{e})}{\|\bar{c}_i\|}.
\end{align*}
Since $c_i$ is $s$-coherent, $\mathrm{sim}(e,\hat{e})\ge s$ for all $\hat{e}\in c_i$, hence
\begin{align*}
\mathrm{sim}(e,\bar{c}_i) & \ge
\frac{1}{|c_i|}\sum_{\hat{e}\in c_i}\frac{\|\hat{e}\|\cdot s}{\|\bar{c}_i\|}
= s\cdot \frac{\frac{1}{|c_i|}\sum_{\hat{e}\in c_i}\|\hat{e}\|}{\|\bar{c}_i\|}\ge s,
\end{align*}
where the last inequality follows from 
\begin{align*}
\frac{1}{|c_i|}\sum_{\hat{e}\in c_i}\|\hat{e}\|\ge
\left\|\frac{1}{|c_i|}\sum_{\hat{e}\in c_i}\hat{e}\right\|=\|\bar{c}_i\|.
\end{align*}
\end{proof}

\vspace{-5px}
\begin{theorem}[Query-to-centroid bound] \label{theo:query_cen_bound_apdx}
Let $u$ be a user-query embedding and $e$ be a relevant knowledge-piece embedding such that $\mathrm{sim}(u,e)\ge k$ for some $k>0$.
If $e$ belongs to an $s$-coherent cluster with centroid $\bar{c}$, then
\[
\mathrm{sim}(u,\bar{c}) \ge ks - \sqrt{(1-k^2)(1-s^2)}.
\]
\end{theorem}
\begin{proof}
From \cref{theo:embed_cen_sim_apdx}, $\mathrm{sim}(e,\bar{c})\ge s$. Let $\alpha$ be the angle between $e$ and $\bar{c}$ and $\beta$ be the angle between $u$ and $e$.
Then $\cos\alpha\ge s$ and $\cos\beta\ge k$, so $\sin\alpha\le \sqrt{1-s^2}$ and $\sin\beta\le \sqrt{1-k^2}$.
The angle between $u$ and $\bar{c}$ is $\alpha+\beta$, hence
\begin{align*}
\mathrm{sim}(u,\bar{c})&=\cos(\alpha+\beta) =\cos\alpha\cos\beta-\sin\alpha\sin\beta 
 \ge ks-\sqrt{(1-k^2)(1-s^2)}.
\end{align*}
\end{proof}

\vspace{-5px}
\paragraph{Conclusion.}
Theorems~\ref{theo:embed_cen_sim_apdx} and \ref{theo:query_cen_bound_apdx} imply that centroid-based routing is \emph{bounded} under the $s$-coherence assumption: if a query is close to a relevant knowledge piece (large~$k$) and clusters are coherent (large~$s$), then the query cannot be arbitrarily far from the corresponding centroid. Therefore, the following failure mode cannot occur in an unbounded way: "a highly specific user query may be very close to the most relevant single knowledge piece but relatively far from the centroid vector of the entire cluster to which the knowledge piece belongs." In practice, centroids can still be less similar than the best individual knowledge piece (because they average multiple pieces), but the gap is bounded by the expression in \cref{theo:query_cen_bound_apdx}.

\begin{table*}[t]
  \centering
  \renewcommand{\arraystretch}{1.05}
  \caption{Performance under different clustering strategies on the NQ dataset (TopK$=5$). Threshold-based clustering and square-root clustering achieve marginally different accuracy, but threshold-based clustering requires the server to store substantially more clusters per agent, which is less practical.}
  \vspace{-8px}
  \resizebox{\linewidth}{!}{
  \begin{tabular}{lccccccccc}
    \Xhline{1.5pt}
    \multirow{2}{*}{\makecell[l]{Clustering strategy}} &
    \multirow{2}{*}{\makecell[c]{Avg Sel. Agents\\(Eff. Rate)}} &
    \multirow{2}{*}{\makecell[c]{Agent-Level\\Ans. Rate}} &
    \multirow{2}{*}{\makecell[c]{\#Clusters\\Per Agent}} &
    \multirow{2}{*}{\makecell[c]{Min/Max\\\#Clusters}} &
    \multicolumn{2}{c}{LLaMA-3.1-8B} && \multicolumn{2}{c}{Qwen-Plus} \\\cline{6-7} \cline{9-10}
    & & & & &  LM & Acc. && LM & Acc. \\\hline
    Threshold-based ($s=0.7$)
    & 3.08 (39.61\%) & 75.13\% & 113.97 & 11 / 267 & 50.17\% & 63.34\% && 61.83\% & 79.86\% \\
    Square-root ($n=\lfloor\sqrt{m}\rfloor$)
    & 3.32 (37.47\%) & 76.47\% & 73.19 & 4 / 214 & 50.98\% & 63.96\% && 62.05\% & 80.05\% \\
    \Xhline{1.5pt}
  \end{tabular}
  }
  \vspace{-15px}
  \label{tab:cluster_threshold_vs_sqrt}
\end{table*}

\vspace{-5px}
\paragraph{Practical design choices: TopK selection and square-root clusters.}
While the threshold-based setup enables analysis, production deployment requires controllable routing cost and stable latency. In \algo, we therefore adopt two practical choices.

First, we select candidate clusters (and thus RAG agents) by TopK, as described in \cref{subsec:online}. A similarity-threshold rule (select all centroids with similarity greater than a designated threshold) is difficult to control: broad queries can match many centroids, leading to unpredictable fan-out and unstable latency/cost. Although \cref{theo:query_cen_bound_apdx} relates cluster coherence to query-centroid similarity, any analytically "safe" threshold can be overly permissive in realistic regimes, still admitting many weakly related clusters. In contrast, TopK is a direct, monotone cost knob and is commonly used to achieve controllable latency and budget constraints. 

Second, as mentioned in \cref{subsec:clustering}, we set the number of clusters per agent to $n=\lfloor\sqrt{m}\rfloor$. This heuristic is inspired by the success of inverted file indexing in database practice: it provides a predictable, sublinear growth of the number of centroids, limiting server-side storage and similarity-search cost. Intuitively, agents with more knowledge pieces may have more overlapping or redundant content and thus can be represented with more clusters, while agents with fewer knowledge pieces are often more specialized and do not require fine-grained partitioning.
This design choice is also supported empirically: \cref{tab:cluster_threshold_vs_sqrt} shows that square-root clustering achieves comparable accuracy to threshold-based clustering while requiring substantially fewer clusters per agent, reducing server-side index/storage overhead.

A similarity threshold can also be used to determine the number of clusters, but it is difficult to calibrate: a high threshold may produce many small clusters, inflating the number of uploaded centroids and thus the server's index size. Moreover, under global TopK selection, an excessive number of clusters per agent can reduce cross-agent coverage: multiple centroids from the same agent may occupy the TopK list, decreasing the number of \emph{unique} selected agents and harming performance on cross-domain queries. For these reasons, we adopt the square-root heuristic and TopK selection as a production-friendly default.

\vspace{-7px}
\subsection{Server Modules} \label{apdx:server_modules}
\vspace{-3px}

\algo coordinates multiple RAG agents using a central server. Beyond routing, the server relies on several \emph{server modules} to (i) filter unreliable or irrelevant responses, (ii) consolidate evidence-backed content, and (iii) refine queries when multi-hop reasoning is required. These server modules do not maintain private knowledge bases; they operate only on the user query, the routed RAG agent responses, and the quoted evidence returned by those RAG agents. Prompt templates are provided in \cref{fig:evaluator_prompts,fig:summarizer_prompts,fig:simplifier_prompts}.

\vspace{-5px}
\paragraph{Response Evaluator (Prompt in \cref{fig:evaluator_prompts}).}
The Response Evaluator is the first checkpoint after collecting responses from selected RAG agents. Since the server lacks domain-specific knowledge, the evaluator uses a common-sense and logic-based assessment with three criteria:
(i) \emph{Relevance}: whether the response helps answer the query (even if incomplete);
(ii) \emph{Evidence-based support}: whether key claims are supported by explicitly quoted evidence; and
(iii) \emph{Logical coherence}: whether the response is consistent and free of contradictions.
For single-hop settings, the evaluator labels each response as \textit{addressed} or \textit{not addressed}.
For multi-hop settings, it further distinguishes \textit{fully addressed} from \textit{partially addressed} among the \textit{addressed} responses.
This filtering is critical because irrelevant RAG agents may still produce plausible but misleading responses; propagating such responses can degrade summarization and cause error accumulation across rounds.

\vspace{-5px}
\paragraph{Response Summarizer (Prompt in \cref{fig:summarizer_prompts}).}
The Response Summarizer consolidates evaluator-approved responses into either a final answer or an intermediate solution.
When at least one response is labeled \textit{fully addressed}, the summarizer produces the final answer by integrating evidence-backed content from one or more fully addressed responses.
When no response is fully addressed but some are \textit{partially addressed}, the summarizer first synthesizes a unified intermediate solution from the valid partial evidence and then re-evaluates whether the unified solution fully resolves the query. If it is deemed fully addressed, the summarizer outputs the final answer; otherwise, it outputs the intermediate solution for query refinement. This design benefits from evaluator filtering, since summarization is more reliable when the input is shorter and contains fewer irrelevant responses.

\vspace{-5px}
\paragraph{Question Simplifier (Prompt in \cref{fig:simplifier_prompts}).}
When the current query is not fully resolved, the Question Simplifier refines it for the next round. Given the original query and the current intermediate solution, it:
(i) identifies which components of the query have already been addressed (with evidence),
(ii) enumerates unresolved aspects that remain necessary for correctness, and
(iii) generates a refined query that removes resolved parts and focuses on unresolved aspects.
This on-the-fly refinement converts partial progress into a targeted next query, enabling iterative multi-hop progress without requiring a rigid precomputed question decomposition plan.

\vspace{-5px}
\subsection{Step-by-Step Procedure for Iterative Refining-Solving} \label{apdx:iterative_procedure}

The main body (\cref{subsec:online}) introduces iterative refining-solving at a high level. This subsection presents a step-by-step procedure that clarifies how routing and server modules interact across rounds for multi-hop queries.

\vspace{-5px}
\paragraph{Inputs.}
A user query $q_0$; a server-side centroid index (uploaded by all RAG agents); a TopK value $K$; and a maximum round limit $T$.

\vspace{-5px}
\paragraph{Outputs.}
A final answer with supporting evidence, or an explicit unanswerable outcome within the current system scope.

\vspace{-5px}
\paragraph{Procedure.}
We maintain the current query $q_t$ and an intermediate solution state (e.g., evidence-backed partial progress) across rounds.
\begin{enumerate}[topsep=0pt,itemsep=0pt,parsep=0pt,partopsep=0pt,leftmargin=*]
    \item \textbf{Initialize.} Set $t=0$ and $q_t=q_0$. Initialize an empty progress record to store intermediate solutions and the evidence used so far.
    \item \textbf{Route and collect responses (core of \rtans).}
    \begin{enumerate}[topsep=0pt,itemsep=0pt,parsep=0pt,partopsep=0pt,leftmargin=*]
        \item Embed the current query to obtain $x=\phi(q_t)$.
        \item Retrieve Top$K$ centroids by similarity and map them to a \emph{deduplicated} set of RAG agents (i.e., if multiple selected centroids belong to the same agent, that agent is invited only once).
        \item Forward $q_t$ to the selected RAG agents in parallel.
        \item Each selected RAG agent runs its local RAG pipeline (retrieval $\rightarrow$ reranking $\rightarrow$ generation) and returns an evidence-backed response.
    \end{enumerate}
    \item \textbf{Evaluate responses.} Apply the Response Evaluator to label each response as \textit{fully addressed}, \textit{partially addressed}, or \textit{not addressed}. Discard responses labeled \textit{not addressed}.
    \item \textbf{Terminate or refine.}
    \begin{enumerate}[topsep=0pt,itemsep=0pt,parsep=0pt,partopsep=0pt,leftmargin=*]
        \item \emph{If any response is fully addressed:} invoke the fully addressed summarizer to produce the final answer and terminate.
        \item \emph{Else if there exist partially addressed responses:} invoke the partially addressed summarizer to synthesize an intermediate solution. If the summarizer deems the unified solution answerable, output the final answer and terminate. Otherwise, proceed to Step~5.
        \item \emph{Else (all responses not addressed):} terminate with an unanswerable outcome (see \cref{apdx:unhandleable}).
    \end{enumerate}
    \item \textbf{Refine the query.} Invoke the Question Simplifier using $(q_t, \text{intermediate solution})$ to produce a refined query $q_{t+1}$ that focuses on unresolved aspects. Set $t\leftarrow t+1$ and repeat from Step~2 unless $t\ge T$.
\end{enumerate}

\vspace{-5px}
\paragraph{Operational notes.}
TopK is the primary cost knob for controlling fan-out and latency. The round limit $T$ prevents runaway loops on ambiguous or out-of-scope queries. Deduplicating selected RAG agents is important when an agent uploads many clusters; otherwise, multiple centroids from a single agent may occupy the TopK list and reduce cross-agent coverage.

\vspace{-5px}
\subsection{Discussion: Unhandleable Queries} \label{apdx:unhandleable}

Even with centroid-based routing and iterative refining-solving, some queries remain unanswerable within the system scope or operational constraints. This can occur for several reasons:
(i) the query falls outside the knowledge coverage of all RAG agents;
(ii) the query is ambiguous, underspecified, or noisy (e.g., due to typos), hindering routing and retrieval;
(iii) the relevant domain evolves rapidly, and agents' local knowledge bases become stale; or
(iv) retrieval/evaluation fails under practical constraints (timeouts, round limits), preventing evidence-backed progress.

\vspace{-5px}
\paragraph{Practical handling.}
In practice, systems typically adopt one of two policies.
First, they may return an explicit unanswerable response to the user and request rephrasing or additional context. Second, they may escalate the query to a human expert, especially in high-stakes settings, optionally forwarding the system's intermediate progress and collected evidence to reduce human effort. 

\vspace{-5px}
\paragraph{What we adopt in this work.}
Our handling differs across settings. In \emph{WikiAgents}, a true human-expert escalation is operationally expensive. We therefore approximate this pathway by invoking a \emph{knowledgeable agent} when \algo fails to obtain any addressed responses (or reaches the round limit). For multi-hop queries, the knowledgeable agent is given a simplified query produced by the Question Simplifier when available, since reducing the number of reasoning steps is often beneficial \cite{zhuang2024efficientrag}. In \emph{NewsAgent}, each input query includes a label indicating whether it is solvable. For queries labeled unsolvable (or when no evidence-backed, addressed responses are obtained), we return an explicit unanswerable response to the user rather than attempting a fallback that could increase hallucination risk in time-sensitive domains.

\vspace{-8px}
\section{Implementations and Baselines} \label{apdx:baselines}

\vspace{-5px}
\subsection{System Implementation Details}
Our implementation builds upon the open-source AgentScope project~\cite{gao2024agentscope}. Each agent's knowledge base is processed by LLaMAIndex~\cite{LiuLlamaIndex2022} and stored in ElasticSearch. We use ElasticSearch mixed retrieval (BM25 + dense vectors) for retrieval. For each selected agent, we retrieve 20 candidate chunks and retain the best five chunks for answer generation. Unless otherwise specified, no more than five agents are selected per routing iteration.

The system is deployed in a pseudo-distributed environment using multi-threading to improve throughput. For model serving, we call Qwen-Plus via API~\cite{yang2024qwen2,bai2023qwen} and deploy LLaMA-3.1-8B locally using vLLM~\cite{kwon2023efficient} on four Nvidia A100 GPUs.

\vspace{-5px}
\subsection{Agent Construction and Corpora}
\paragraph{WikiAgents.}
WikiAgents is built over 121K Wikipedia pages (snapshot date: Nov.\ 1, 2023) and contains exactly 64 agents. This number follows the limitation of the ORES legacy service, which assigns Wikipedia pages into 64 predefined categories~\cite{johnson2021classification}; we map each category to one agent so that agent specialization aligns with the ORES taxonomy. Because this taxonomy may not cover all documents required by downstream questions, when retrieved evidence is insufficient, the LLM may still respond based on its parametric knowledge. We account for this behavior in evaluation by reporting both lexical match and LLM-judged accuracy (Section~\ref{subsec:setup}). 

\paragraph{NewsAgents.}
NewsAgents is constructed from 609 news articles from 49 media outlets across six domains~\cite{tang2024multihop}.
We consider two partitions: \textit{NewsAgent-Source} (agents grouped by outlet) and \textit{NewsAgent-Domain} (agents grouped by domain).
Each agent uses LLaMAIndex to split articles into 256-token chunks with 20-token overlap between consecutive chunks.

\vspace{-5px}
\subsection{Baselines}\label{sec:baselines}
We reproduce baselines within our multi-agent evaluation framework. All methods are evaluated in an off-the-shelf setting: we do not train any models, and instead rely on LLM capabilities.

\vspace{-5px}
\paragraph{Single-agent Scenarios.}
In the single-agent setting, the knowledge contained within all RAG agents is merged into a single RAG system. This provides a conventional RAG baseline that can retrieve from the entire corpus. Although it does not fully match the multi-agent routing scenario, it illustrates challenges of large-scale retrieval (e.g., longer retrieval latency and increased distractors). If retrieval were perfectly clean, the single-agent setup could be viewed as an approximate upper bound for multi-agent accuracy when judged against ground truth; in practice, however, pooling all documents increases the chance of retrieving distractors, especially for multi-hop queries. 

\begin{itemize}[topsep=0pt,itemsep=0pt,parsep=0pt,partopsep=0pt,leftmargin=*]
    \item \textbf{EfficientRAG~\cite{zhuang2024efficientrag}:}
    This approach iteratively simplifies/decomposes the query, retrieving evidence at each step until the question becomes answerable.
    The simplification process is performed by the LLM by removing resolved parts of the query; if further progress is not possible, a knowledgeable agent produces the final answer.
    Our implementation follows EfficientRAG~\cite{zhuang2024efficientrag} and is conceptually related to Self-ASK~\cite{press2022measuring}, SelfRAG~\cite{asai2023self}, and IM-RAG~\cite{yang2024rag}.
\end{itemize}

\paragraph{Multi-agent Scenarios.}
In the multi-agent setting, we keep the same agent backends and retrieval modules as \algo, and only replace the routing mechanism.

\begin{itemize}[topsep=0pt,itemsep=0pt,parsep=0pt,partopsep=0pt,leftmargin=*]
    \item \textbf{Chameleon~\cite{lu2024chameleon}:} A description-based router that prompts an LLM with agent descriptions and selects the agents most relevant to the current query. For WikiAgents, each description is derived from the corresponding ORES topic category. For NewsAgents, descriptions are derived from either the news outlet or the domain, depending on the partition.

    \item \textbf{RouterDC~\cite{chen2024routerdc}:} A learning-based router that represents each candidate agent with a learned routing vector and selects agents by embedding similarity at inference time. To instantiate this baseline in our setting, we collect 100 single-hop answerable questions for each WikiAgent and use them to fit agent representations while keeping the embedding backbone fixed. We do not report RouterDC on MHR because that benchmark does not provide enough agent-specific supervision to train these representations reliably.
\end{itemize}

\vspace{-5px}
\section{More Quantitative Results}

\paragraph{Response Evaluator Performance under Different Models.} 
\begin{wraptable}{r}{0.5\textwidth}
\vspace{-10pt}
\centering
\caption{Ablation Study: Response Evaluator Performance under different models on the Natural Questions dataset}
\vspace{-8px}
\renewcommand{\arraystretch}{1.05}
\resizebox{.5\textwidth}{!}{
\begin{tabular}{lcc}
\Xhline{1.5pt}
 & LLaMA-3.1-8B & Qwen-Plus \\
\hline
\multicolumn{3}{@{}l}{\textit{Response-level (per response)}} \\
Correctness & 45.60\% & 79.09\% \\ 
Addressed Correctness & 50.92\% & 88.73\% \\
\hline
\multicolumn{3}{@{}l}{\textit{Query-level (per query)}} \\
Fully Correct & 16.48\% & 54.03\% \\
All Addressed Correct & 16.85\% & 59.98\% \\
\Xhline{1.5pt}
\end{tabular} }

\label{tab:performance}
\end{wraptable}
In this experiment, we evaluate a response evaluator on a single-hop QA dataset (Natural Questions). We construct an evaluation set of queries from Natural Questions, agent-generated responses, and correctness annotations. The correctness labels are provided by GPT-4o, which compares each agent response with the ground-truth answer. Notably, the same query may appear multiple times, each paired with responses from different agents. The response evaluator, implemented with either LLaMA-3.1-8B or Qwen-Plus, classifies each response as \emph{addressed} if it directly answers the query, and \emph{not addressed} otherwise. We report: (i) \textbf{Fully Correct}, the fraction of queries for which all addressed responses are correct and all not-addressed responses are incorrect; (ii) \textbf{All Addressed Correct}, the fraction of queries for which every addressed response is correct; (iii) \textbf{Evaluator Correction}, the probability that the evaluator assigns the correct addressed/not-addressed label; and (iv) \textbf{Evaluator Addressed Correctness}, i.e., the probability that a response labeled as \emph{addressed} is actually correct. 

As shown in Table~\ref{tab:performance}, Qwen-Plus substantially outperforms LLaMA-3.1-8B as a response evaluator at both the response level (Evaluator Correctness and Evaluator Addressed Correctness) and the query level (Fully Correct and All Addressed Correct). This indicates that Qwen-Plus is more reliable at flagging whether an agent response truly addresses the question and, crucially, that its \emph{addressed} predictions are much more likely to be correct. Such stronger filtering reduces error propagation across routing iterations and helps explain the gains of the mixed setting reported in Section~\ref{subsec: single-hop-exp}. 

\begin{table*}[t]
    \centering
    \caption{Accuracy comparison on 2WikiMultiHopQA and HotpotQA.}
    \vspace{-7px}
    \renewcommand{\arraystretch}{1.2}
    \resizebox{\textwidth}{!}{
    \begin{tabular}{l l | c c c c | c c c c}
        \Xhline{2pt}
        \multirow{3}{*}{Methods} & \multirow{3}{*}{Models} 
        & \multicolumn{4}{c|}{2WikiMultiHopQA} 
        & \multicolumn{4}{c}{HotpotQA} \\\cline{3-10}
        & & Acc. & \makecell{Ans. rate\\by RAG} & RAG Acc. & CoT Acc. 
          & Acc. & \makecell{Ans. rate\\by RAG} & RAG Acc. & CoT Acc. \\
        \hline
        
        \multirow{2}{*}{CoT}
            & Qwen-Plus & 37.31 & - & - & 37.31 & 54.58 & - & - & 54.58 \\
            & LLaMA-3.1-8B & 25.99 & - & - & 25.99 & 44.23 & - & - & 44.23 \\
        \hline
        
        \multirow{2}{*}{Chameleon}
            & Qwen-Plus 
                & 60.12 & 68.92 & 72.86 & 31.87
                & 73.19 & \textbf{84.10} & 78.70 & 44.56 \\
            & LLaMA-3.1-8B 
                & 40.27 & 67.81 & 45.89 & 28.41
                & 50.20 & \underline{77.75} & 55.81 & 30.56 \\
        \hline
        
        \multirow{2}{*}{RouterDC}
            & Qwen-Plus 
                & 52.12 & 62.20 & 62.47 & 35.63
                & 67.24 & 72.26 & 75.64 & 45.32 \\
            & LLaMA-3.1-8B 
                & 30.59 & 68.33 & 34.25 & 22.69
                & 47.74 & 74.51 & 50.92 & 38.51 \\
        \hline
        
        \multirow{3}{*}{\algo}
            & Qwen-Plus 
                & \textbf{65.78} & \textbf{78.98} & \textbf{74.68} & 31.88
                & \textbf{75.63} & 82.85 & \textbf{81.86} & 45.37 \\
            & LLaMA-3.1-8B 
                & \underline{45.65} & \underline{72.85} & \underline{51.62} & 29.60
                & \underline{58.69} & 77.33 & \underline{64.42} & 39.23 \\
            & Mixed 
                & 62.06 & 76.93 & 69.65 & 36.74
                & 69.33 & 80.27 & 73.17 & 53.70 \\
        \Xhline{2pt}
    \end{tabular}}
    \vspace{-15px}
    \label{tab:wiki_comparison_revised}
\end{table*}

\begin{figure*}[t]
    \centering
    \subfloat[MuSiQue]{%
        \includegraphics[width=0.32\linewidth]{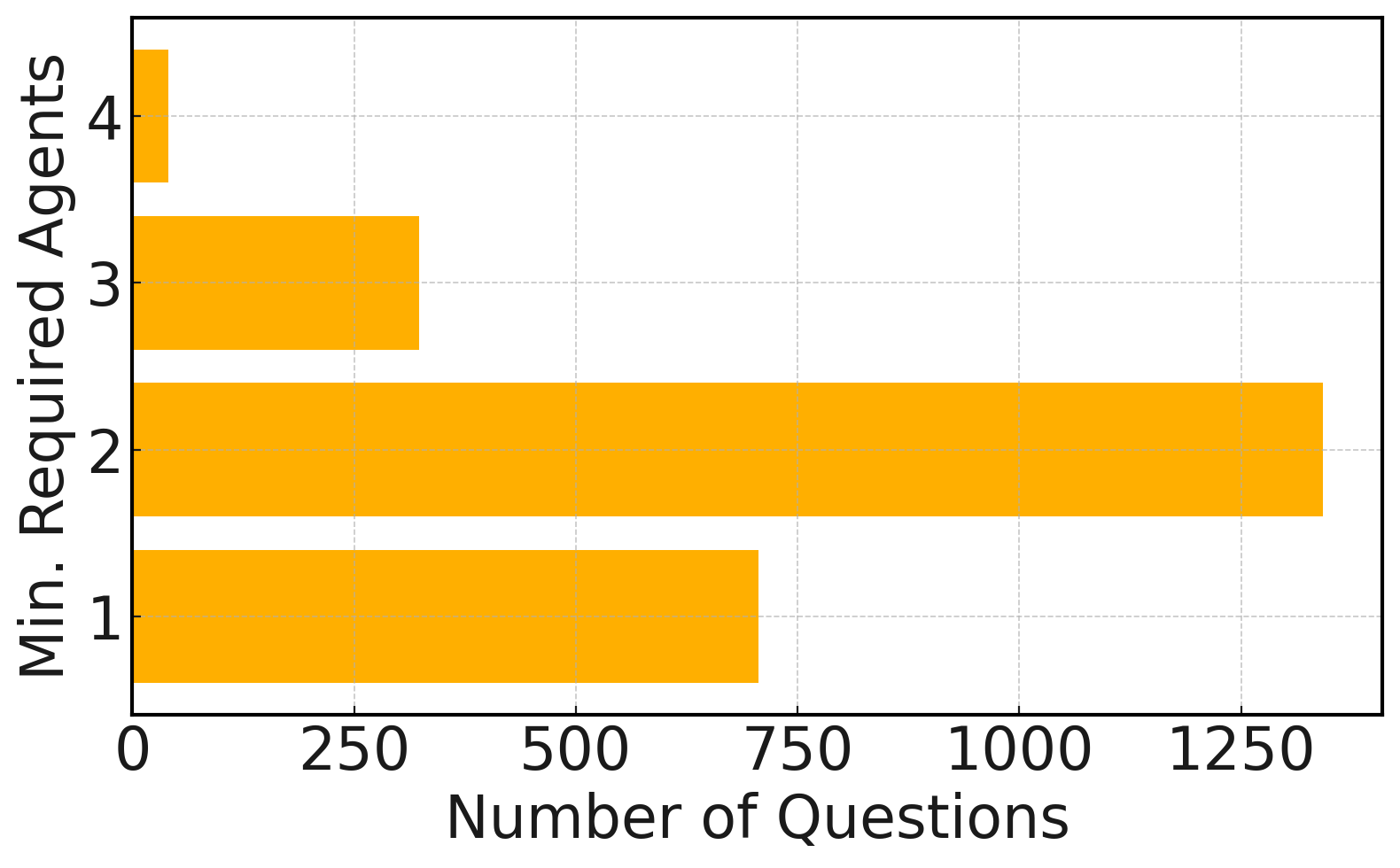}
    }
    \hfill
    \subfloat[2WikiMultiHopQA]{%
        \includegraphics[width=0.32\linewidth]{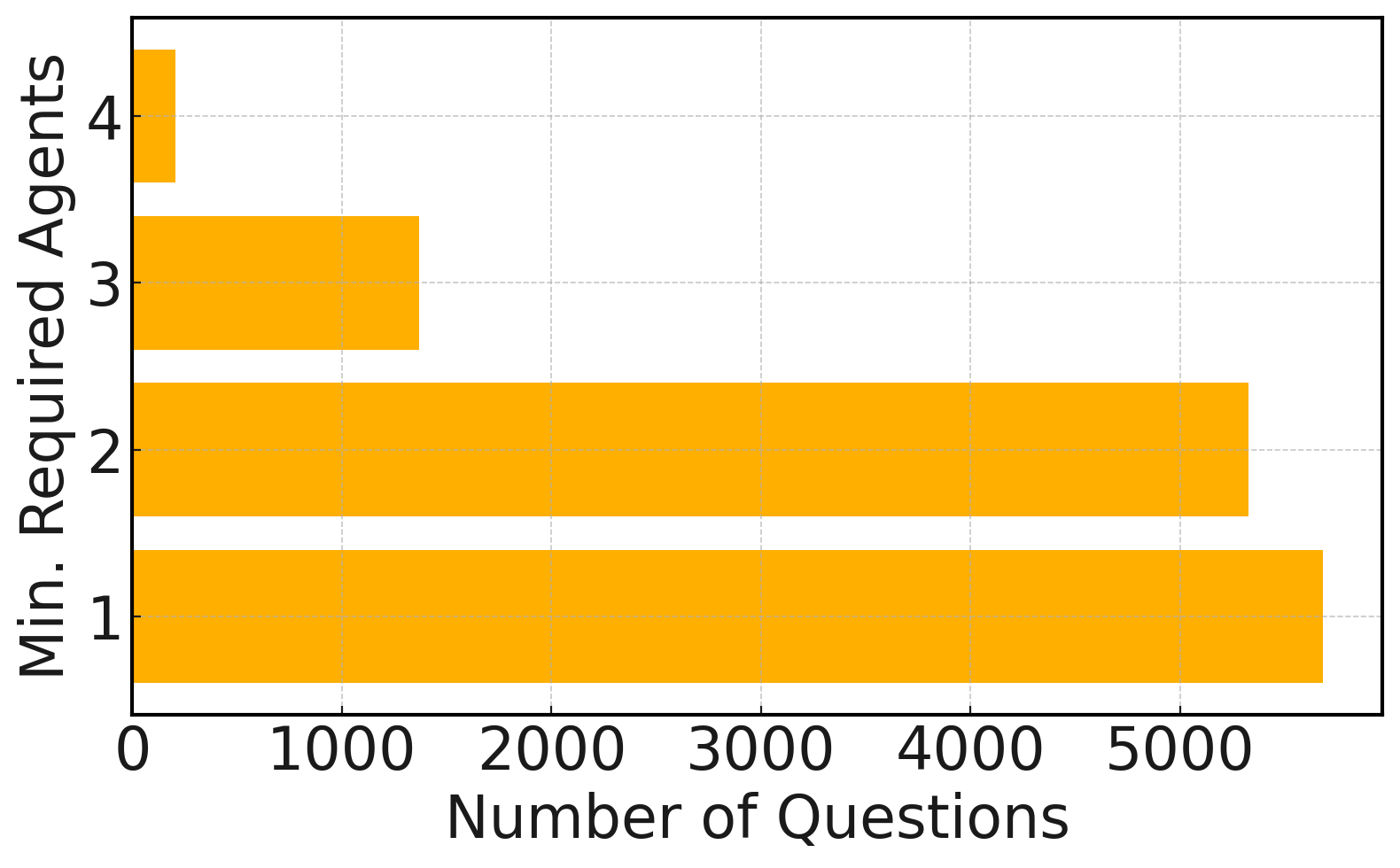}
    }
    \hfill
    \subfloat[HotpotQA]{%
        \includegraphics[width=0.32\linewidth]{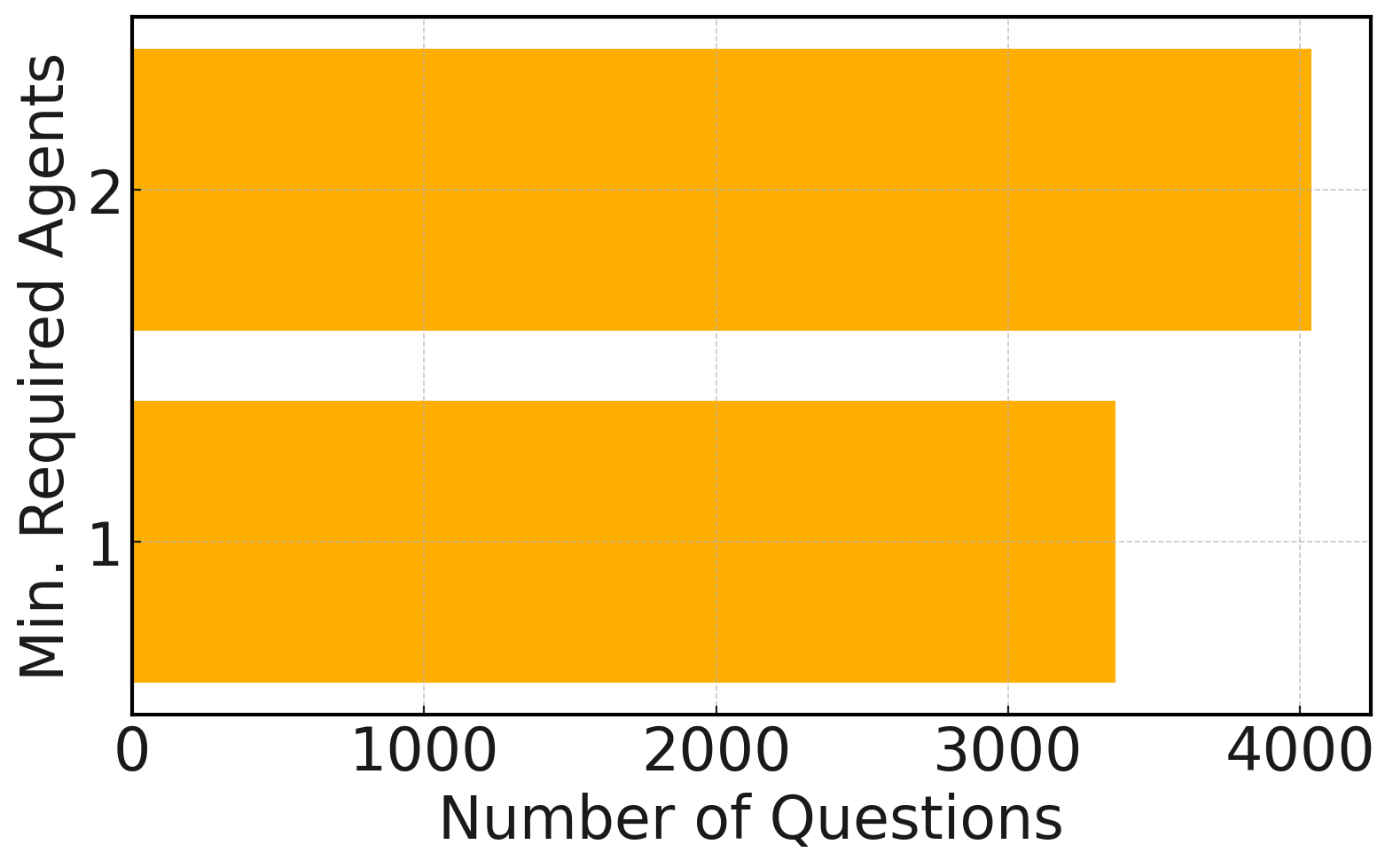}
    }
    \vspace{-10px}
    \caption{Data distribution v.s. Minimum required agents under different datasets}
    \vspace{-15px}
    \label{fig:wikiqa_dist}
\end{figure*}

\begin{figure*}[htbp]
    \centering
    \subfloat[Qwen-Plus]{%
        \includegraphics[width=0.32\linewidth]{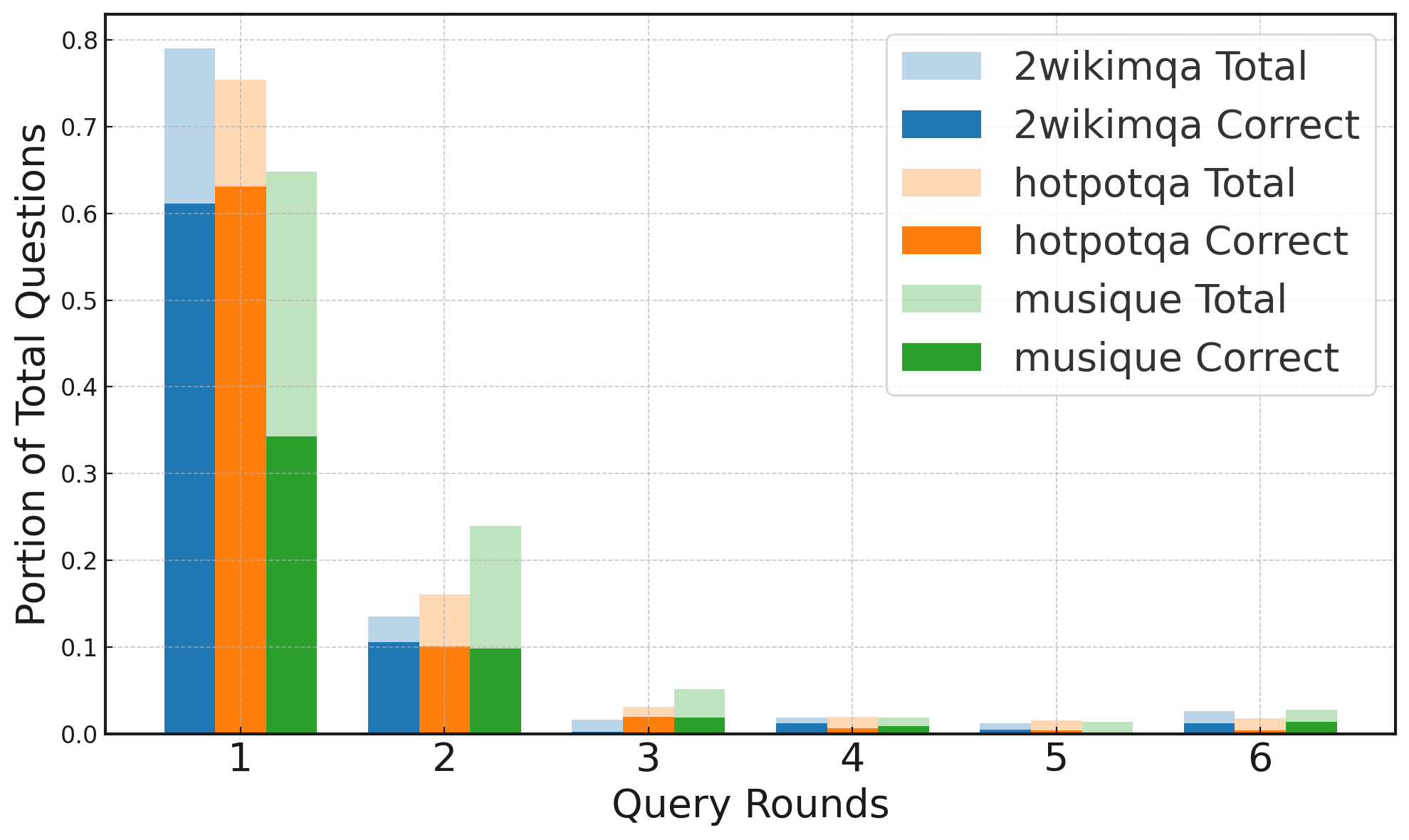}
    }
    \hfill
    \subfloat[LLaMA-3.1-8B]{%
        \includegraphics[width=0.32\linewidth]{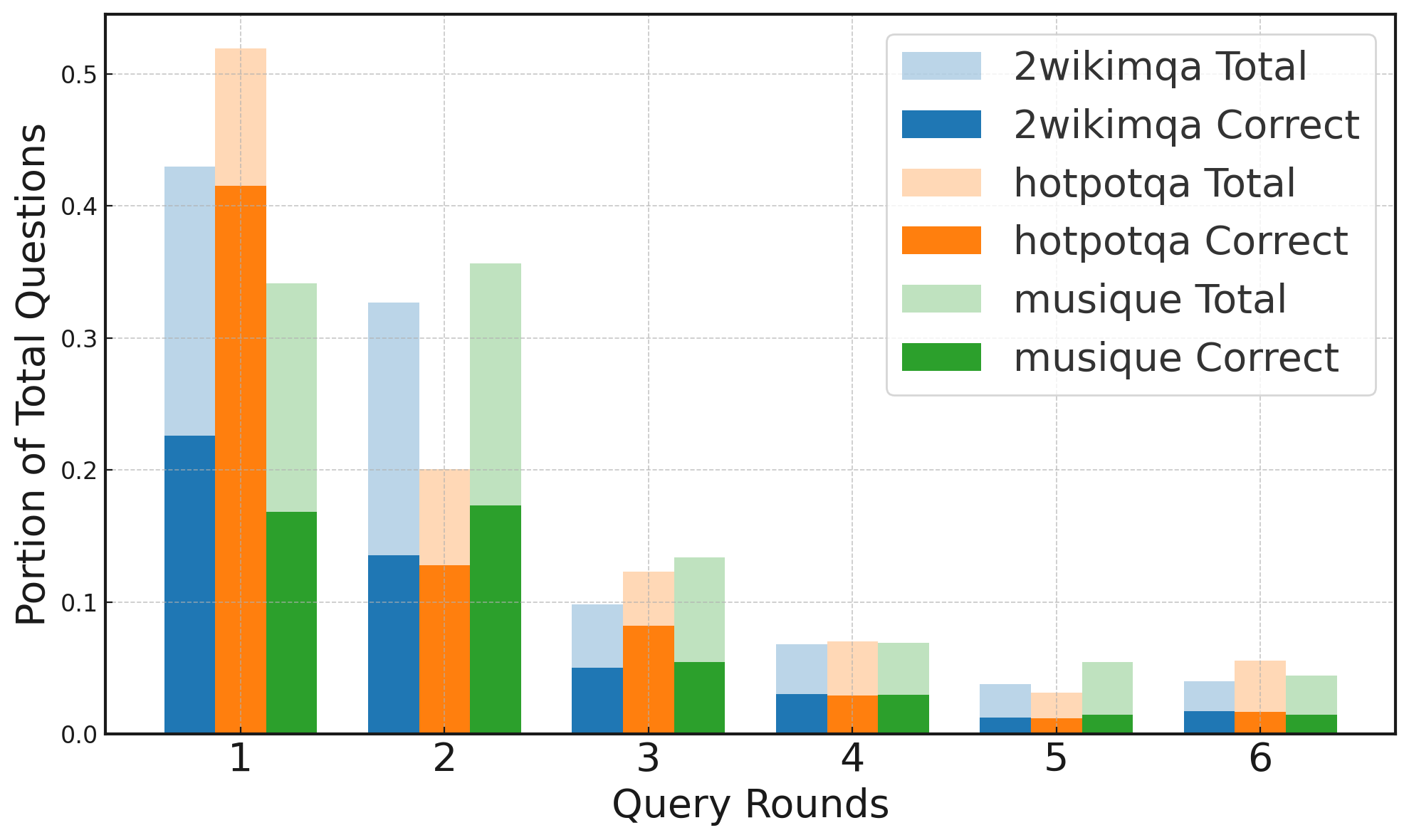}
    }
    \hfill
    \subfloat[Mixed]{%
        \includegraphics[width=0.32\linewidth]{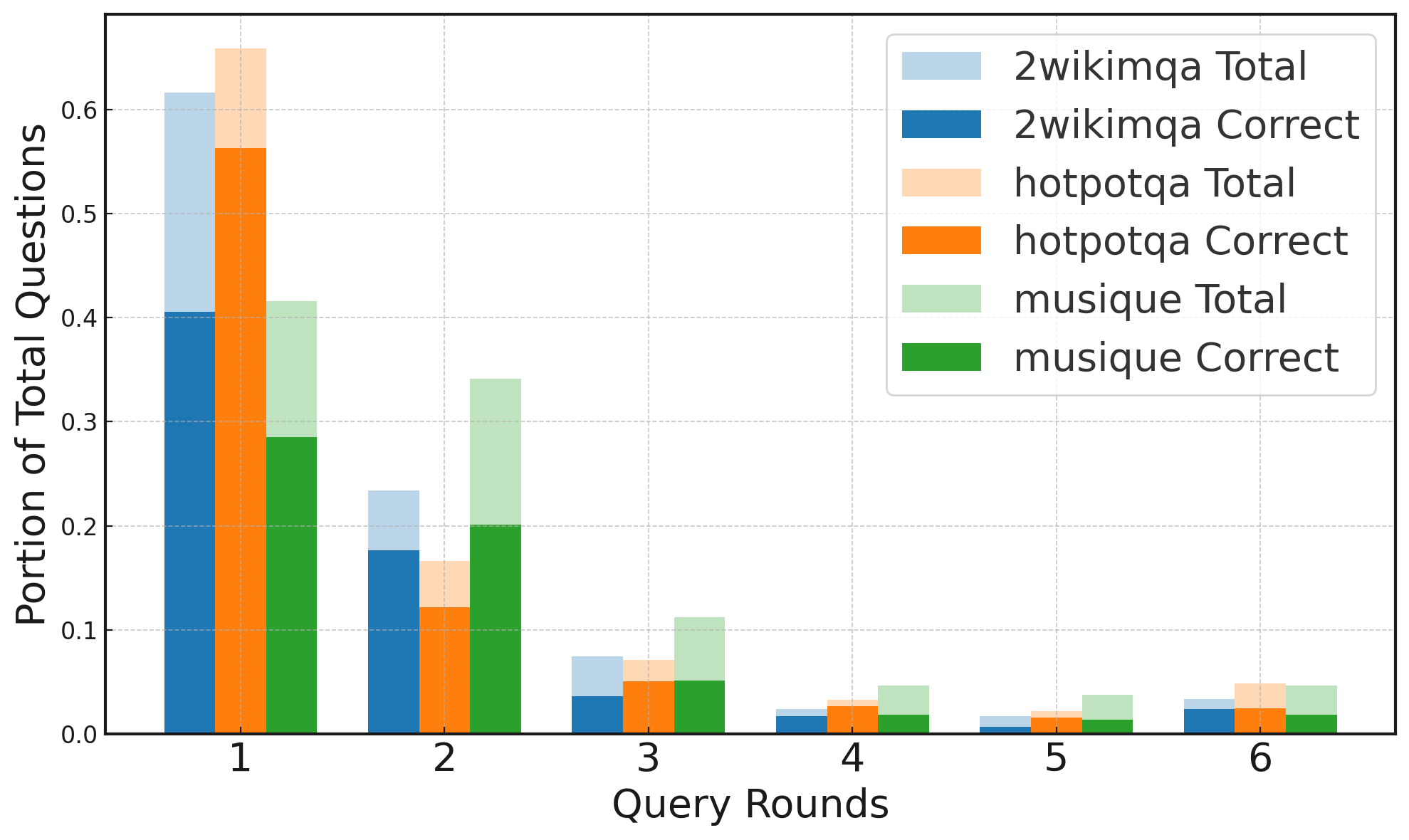}
    }
    \vspace{-10px}
    \caption{Evaluation for Wikipedia-related QA (2WikiMQA, HotpotQA, and MuSiQue)}
    \vspace{-10px}
    \label{fig:wikiqa_eval}
\end{figure*}

\vspace{-5px}
\paragraph{Performance when Excluding Knowledgeable Agents.} 
\cref{tab:wiki_comparison_revised} compares Chameleon, RouterDC, and \algo on 2WikiMultiHopQA and HotpotQA. Across both datasets, \algo achieves the highest answerable rate according to RAG and the highest RAG accuracy, indicating more reliable agent selection and stronger evidence-grounded generation. Chameleon generally achieves higher RAG accuracy than RouterDC, while \algo improves on both dimensions, which helps explain its overall accuracy gains. The CoT accuracy of these retrieval-based methods can be either higher or lower than the CoT-only baseline, with no deterministic trend: if many RAG-unanswered questions are still solvable via the model's parametric knowledge, CoT accuracy can increase; conversely, if the remaining unanswered questions are intrinsically difficult or poorly handled without evidence, CoT accuracy can decrease. 

\vspace{-5px}
\paragraph{Number of Required Agents.} 
Figure \ref{fig:wikiqa_dist} illustrates the distribution of questions by the minimum number of required agents across three multi-hop QA datasets: MuSiQue, 2WikiMultiHopQA, and HotpotQA. Although all three datasets consist of multi-hop questions, the necessary knowledge pieces may sometimes reside within a single agent. In MuSiQue and 2WikiMultiHopQA, most questions can be answered using two RAG agents, and all questions require no more than four agents. For HotpotQA, all questions can be answered with just two agents. It is important to note that the number of required agents does not directly correspond to the number of query rounds. Multiple agents can be invoked simultaneously within a single round, and conversely, a single agent may be queried multiple times to perform complex reasoning before reaching the final answer.

\vspace{-5px}
\paragraph{Query Rounds and Accuracy.} 
Figure \ref{fig:wikiqa_eval} presents the proportion of questions that can be answered within different numbers of query rounds, along with the corresponding answer correctness. Across all LLMs, most questions are resolved within the first two query rounds. Specifically, stronger models (e.g., Qwen-Plus) can answer more questions compared to weaker ones (e.g., LLaMA-3.1-8B) within fewer query rounds. Interestingly, the figure also shows that increasing the number of query rounds does not necessarily lead to improved answer accuracy. In fact, unless a question inherently requires multi-step reasoning, additional query rounds may result in even poorer performance. 

This trend is consistent with the routing-error analysis in Section~\ref{subsec: single-hop-exp}: additional rounds create more opportunities for misrouting and for accumulating partially relevant evidence, which can distract the generator. Therefore, accurate answerability judgment and effective stopping criteria are crucial for controlling both latency and error accumulation.

\vspace{-5px}
\paragraph{Case Study.}
Figures \ref{fig:sample_single_qa} and \ref{fig:sample_multihop_qa} illustrate how the proposed \algo answers a single-hop question and a multi-hop question, respectively.

\newpage


\newcommand{\myparatight}[1]{\smallskip\noindent{\bf {#1}.}~}
\tcbset{
    userstyle/.style={
        enhanced,
        colback=white,
        colframe=black,
        colbacktitle=gray!20,
        coltitle=black,
        rounded corners,
        sharp corners=north,
        boxrule=0.5pt,
        drop shadow=black!50!white,
        attach boxed title to top left={
            xshift=-2mm,
            yshift=-2mm
        },
        boxed title style={
            rounded corners,
            size=small,
            colback=gray!20
        },
        fontupper=\footnotesize,
        left=1mm,
        right=1mm,
        top=2mm,
        bottom=1mm
    },
    jailbreakstyle/.style={
        enhanced,
        colback=white,
        colframe=red,
        colbacktitle=red!40,
        coltitle=black,
        rounded corners,
        sharp corners=north,
        boxrule=0.5pt,
        drop shadow=red!50!white,
        attach boxed title to top left={
            xshift=-2mm,
            yshift=-2mm
        },
        boxed title style={
            rounded corners,
            size=small,
            colback=red!20
        },
        fontupper=\footnotesize,
        left=1mm,
        right=1mm,
        top=2mm,
        bottom=1mm
    },
    jailbreakstyleres/.style={
        enhanced,
        colback=white,
        colframe=red,
        colbacktitle=red!40,
        coltitle=black,
        rounded corners,
        sharp corners=north,
        boxrule=0.5pt,
        drop shadow=red!50!white,
        attach boxed title to top right={
            xshift=-2mm,
            yshift=-2mm
        },
        boxed title style={
            rounded corners, 
            size=small,
            colback=red!0
        },
        fontupper=\footnotesize,
        left=1mm,
        right=1mm,
        top=2mm,
        bottom=1mm
    },
    myreplyborderstyle/.style={
        enhanced,
        colback=white,
        colframe=black,
        colbacktitle=red!40,
        coltitle=black,
        rounded corners,
        sharp corners=north,
        boxrule=0.5pt,
        drop shadow=black!50!white,
        attach boxed title to top right={
            xshift=-2mm,
            yshift=-2mm
        },
        boxed title style={
            rounded corners, 
            size=small,
            colback=red!0
        },
        fontupper=\footnotesize,
        left=1mm,
        right=1mm,
        top=2mm,
        bottom=1mm
    },
    replystyleg/.style={
        enhanced,
        colback=blue!0,
        colbacktitle=black,
        colframe=black,
        coltitle=black,
        boxrule=1pt,
        drop shadow=black!50!,
        rounded corners,
        sharp corners=north,
        attach boxed title to top right={
            xshift=-2mm,
            yshift=-2mm
        },
        boxed title style={
            rounded corners,
            size=small, 
            colback=blue!0,
        },
        fontupper=\footnotesize,
        left=1mm,
        right=1mm,
        top=2mm,
        bottom=1mm
    },
    replystyler/.style={
        enhanced,
        colback=pink!10,
        colframe=black,
        colbacktitle=red!15,
        coltitle=black,
        boxrule=0.5pt,
        drop shadow=black!50!white,
        rounded corners,
        sharp corners=north,
        attach boxed title to top right={
            xshift=-2mm,
            yshift=-2mm
        },
        boxed title style={
            rounded corners,
            size=small,
        },
        fontupper=\footnotesize,
        left=1mm,
        right=1mm,
        top=2mm,
        bottom=1mm
    },
    replystylew/.style={
        enhanced,
        colback=cyan!5,
        colframe=black,
        colbacktitle=green!40,
        coltitle=black,
        boxrule=0.5pt,
        drop shadow=black!50!white,
        rounded corners,
        sharp corners=north,
        attach boxed title to top right={
            xshift=-2mm,
            yshift=-2mm
        },
        boxed title style={
            rounded corners,
            size=small,
            colback=cyan!20
        },
        fontupper=\footnotesize,
        left=1mm,
        right=1mm,
        top=2mm,
        bottom=1mm
    },
    replystyleo/.style={
        enhanced,
        colback=orange!5,
        colframe=black,
        colbacktitle=green!40,
        coltitle=black,
        boxrule=0.5pt,
        drop shadow=black!50!white,
        rounded corners,
        sharp corners=north,
        attach boxed title to top right={
            xshift=-2mm,
            yshift=-2mm
        },
        boxed title style={
            rounded corners,
            size=small,
            colback=orange!20
        },
        fontupper=\footnotesize,
        left=1mm,
        right=1mm,
        top=2mm,
        bottom=1mm
    },
    replystylegr/.style={
        enhanced,
        colback=green!5,
        colframe=black,
        colbacktitle=green!40,
        coltitle=black,
        boxrule=0.5pt,
        drop shadow=black!50!white,
        rounded corners,
        sharp corners=north,
        attach boxed title to top right={
            xshift=-2mm,
            yshift=-2mm
        },
        boxed title style={
            rounded corners,
            size=small,
            colback=green!20
        },
        fontupper=\footnotesize,
        left=1mm,
        right=1mm,
        top=2mm,
        bottom=1mm
    }
}

\newtcolorbox{userquery}[1][]{
    userstyle,
    title=Prompt,
    #1
}

\newtcolorbox{llmreply-g}[1][]{
    replystyleg,
    title=Response,
    #1
}

\newtcolorbox{llmreply-r}[1][]{
    replystyler,
    title=Response,
    #1
}

\newtcolorbox{mybox}[2][]{
    replystyler,
    title=#2,
    #1
}
\newtcolorbox{myboxw}[2][]{
    replystylew,
    title=#2,
    #1
}

\newtcolorbox{myboxo}[2][]{
    replystyleo,
    title=#2,
    #1
}

\newtcolorbox{myboxg}[2][]{
    replystyler,
    title=#2,
    #1
}

\newtcolorbox{myboxgr}[2][]{
    replystylegr,
    title=#2,
    #1
}

\newtcolorbox{myuser}[2][]{
    userstyle,
    title=#2,
    #1
}

\newtcolorbox{myjailbreak}[2][]{
    jailbreakstyle,
    title=#2,
    #1
}

\newtcolorbox{myreplyborder}[2][]{
    myreplyborderstyle,
    title=#2,
    #1
}

\renewcommand{\paragraph}[1]{\noindent\textbf{#1~}}

\begin{figure*}[h]
    \centering 
    \hspace{-2mm}
\begin{myuser}{System Prompts for Chain-of-Thought (CoT) Agent}
You're a knowledgeable assistant. You are provided with a question, and you should answer the question in the following two steps. FIRST, you should utilize your knowledge and analyze the question step by step. SECOND, you should finalize an answer based on your analysis with no more than 30 words. \newline
\newline
Your output should be in the json format:\newline
\verb|```|\texttt{json}
\begin{lstlisting}[language=json]
{
    "analysis": "<a paragraph with no more than six sentences>",
    "answer": "<a response within 30 words>"
}
\end{lstlisting}
\verb|```|
\end{myuser}

    \vspace{-2mm} 
    \caption{Prompt Templates for the CoT Agents.} 
    \label{fig:cot_agent_prompts} 
\end{figure*}

\begin{figure*}[h]
    \centering 
    \hspace{-2mm}
\begin{myuser}{System Prompts of Wikipedia Agents for A Single Question}
You are provided with one question and a collection of knowledge. Based on the given knowledge, you should try to analyze and tackle the question as thoroughly as possible, even if you cannot fully answer the question. Unless the given context is unrelated to the question, you must directly quote the evidence (i.e., sentences) without being altered to support your analysis, enclosing it in double asterisks (**). You should not state any arguments that are not explicitly mentioned or implied from the pieces of evidence or without quoting them. The analysis should be in one paragraph with no more than ten sentences. Moreover, the analysis should start with "I" and not mention that the analysis is generated based on the given knowledge, documents, or information.\newline\newline
Your output should be in the json format:\newline
\verb|```|\texttt{json}
\begin{lstlisting}[language=json]
{
    "analysis": "<one paragraph of up to ten sentences, directly quoting supporting evidence from the provided knowledge>"
}
\end{lstlisting}
\verb|```|
\end{myuser}

    \vspace{-2mm} 
    \caption{Prompt Templates for the Wikipedia Agents.} 
    \vspace{-5mm}
    \label{fig:wiki_agent_prompts} 
\end{figure*}

\begin{figure*}[h]
    \centering 
    \hspace{-2mm}
\begin{myuser}{System Prompts of News Agents for A Single Question}
You are given one question and a collection of news articles. Each article contains content along with its source information, including the title, news source, author, and published time. Your task is to analyze and address the question as thoroughly as possible based on the provided news articles, even if you cannot fully answer the question. Unless the given context is unrelated to the question, you must directly quote the evidence (i.e., sentences) without being altered to support your analysis. You should enclose quoted evidence (sentences) in double asterisks (**), followed by the source in brackets, including the title, news source, author, and published time, separated by semicolons (;). You should not state any arguments that are not explicitly mentioned or implied from the pieces of evidence or without quoting them. Your analysis to each question should be concise, limited to one paragraph per question, with no more than ten sentences. The analysis must begin with "I" and should not mention that the analysis is based on provided news articles, knowledge, or information.\newline
\newline
Your output should be in the json format:\newline
\verb|```|\texttt{json}
\begin{lstlisting}[language=json]
{
    "analysis": "<one paragraph of up to ten sentences, directly quoting supporting evidence from the provided knowledge>"
}
\end{lstlisting}
\verb|```|
\end{myuser}
    \vspace{-2mm} 
    \caption{Prompt Templates for the News Agents.} 
    \label{fig:mhr_agent_prompts} 
\end{figure*}

\begin{figure*}[h]
    \centering 
    \hspace{-2mm}
\begin{myuser}{System Prompts of Evaluator for "Fully/Partially/Not Addressed" for A Question and A Response}
You are provided with a question and a response. Your task is to evaluate the response according to the following steps:\newline
1. Assess the response against three criteria:\newline
    - Relevance: Does the response help answer the question, even if the response does not fully resolve it?\newline
    - Evidence-Based Support: Are statements supported by explicitly mentioned evidence enclosed in double asterisks (**)?\newline
    - Logical Coherence: Is it well-structured, logically reasoned, and free from logical fallacies or contradictions?\newline
2. Assign one of the following ratings:\newline
    - Fully addressed: The response meets all criteria and completely answer the question.\newline
    - Partially addressed: The response meets all criteria but not fully resolve the question.\newline
    - Not addressed: The response fails to meet one or more of the criteria.\newline
\newline
Your output should be in the json format:\newline
\verb|```|\texttt{json}
\begin{lstlisting}[language=json]
{
    "evaluation": {
        "relevance": <a sentence within 30 words>, 
        "evidence_support": <a sentence within 30 words>, 
        "logical_coherence": <a sentence within 30 words>
    }, 
    "rating": <"Fully addressed" or "Partially addressed" or "Not addressed">
}
\end{lstlisting}
\verb|```|
\end{myuser}
    \vspace{-2mm} 
    \caption{Prompt Templates for the Response Evaluator.} 
    \vspace{-5mm}
    \label{fig:evaluator_prompts} 
\end{figure*}

\begin{figure*}[h]
    \centering 
    \hspace{-2mm}
\begin{myuser}{System Prompts of Summarizing Fully-addressed Responses}
You are given a question, and one or more responses that fully resolve the question. Your task is to produce a final answer by following these steps:\newline
1. Analysis: Incorporate all relevant information from the given responses, quoting any supporting evidence word-for-word in double asterisks (**).\newline
2. Answer: Provide a concise conclusion in no more than 30 words that summarizes the analysis.\newline
\newline
Your output should be in the json format:\newline
\verb|```|\texttt{json}
\begin{lstlisting}[language=json]
{
    "analysis": "<a paragraph that directly quotes relevant evidence in **double asterisks**>", 
    "answer": "<a concise final answer within 30 words>"
}
\end{lstlisting}
\verb|```|
\end{myuser}

\begin{myuser}{System Prompts of Summarizing Partially-addressed Responses}
You are given a question and several partially addressed responses. Your task is to combine these responses to create a comprehensive solution, then evaluate its completeness. Follow these steps:\newline
1. Solution Synthesis:\newline
    - Incorporate all relevant information from the provided responses to form a solution that addresses the question as thoroughly as possible.\newline
    - Directly quote supporting evidence (word-for-word) using double asterisks (**).\newline
2. Evaluation and Justification:\newline
    - Determine whether this combined solution fully addresses the question.\newline
    - Provide a clear explanation of why it does or does not fully address the question.\newline
3. Answerability Determination:\newline
    - Based on your evaluation, decide if the solution makes the question answerable.\newline
    - Respond with either "yes" (if the solution fully addresses the question) or "no" (if it does not).\newline
    - Most importantly, if the solution explicitly states that the question cannot be fully addressed or identifies missing aspects or necessary additional information, you must answer "no".\newline
4. Final Answer:\newline
    - If the solution is deemed answerable ("yes"), provide a concise conclusion in no more than 30 words that summarizes the solution.\newline
    - If the solution is not answerable ("no"), return "None" as the final answer.\newline
\newline
Your output should be in the json format:\newline
\verb|```|\texttt{json}
\begin{lstlisting}[language=json]
{
    "solution": "<a paragraph that integrates responses with direct quotes in **double asterisks**>",
    "evaluation and justification": "<a paragraph explaining whether and why the solution fully addresses the question>",
    "answerable": <"yes" or "no">,
    "answer": "<a final answer within 30 words if 'answerable' is 'yes', or 'None' if 'answerable' is 'no'>"
}
\end{lstlisting}
\verb|```|
\end{myuser}

    \vspace{-3mm} 
    \caption{Prompt Templates for the Response Summarizer.} 
    \vspace{-5mm}
    \label{fig:summarizer_prompts} 
\end{figure*}   

\begin{figure*}[h]
    \centering 
    \hspace{-2mm}
\begin{myuser}{System Prompts of Question Simplifier}
You are given a question and a piece of knowledge that partially addresses the question. Your task is to simplify or refactor the original question so that answering the simplified question will yield the same final answer as answering the original one. Follow the steps below:\newline
1. Map Known Information:\newline
    - Identify Answered Parts: Examine the original question and determine which parts have been explicitly answered by the provided solution. Extract the relevant words or phrases from the question.\newline
    - Map to Solution: For each identified part, find the corresponding words or phrases in the provided solution and include supporting evidence by directly quoting the exact sentences, enclosed in double asterisks.\newline
    - Note: If the provided knowledge does not address any aspect of the original question (i.e., you cannot identify the answered parts from the question), return "None" for the mapping.\newline
2. Identify Unresolved Aspects:\newline
    - Determine Gaps: Identify the parts of the question that remain unanswered or unclear after considering the provided solution.\newline
    - List Unresolved Points: Clearly list these unresolved aspects. They must be retained in the simplified question to ensure that no essential detail is lost.\newline
3. Generate a Simplified/Refactored Question:\newline
    - Remove Redundancies: Remove from the original question any words or phrases that have been fully addressed by the solution (as determined in Step 1).\newline
    - Incorporate Mapped Knowledge: Replace the removed parts with their corresponding mapped words or phrases from the solution, ensuring to preserve the supporting evidence.\newline
    - Retain Unresolved Aspects: Ensure that any unresolved aspects identified in Step 2 are still present in the final simplified question.\newline
    - Ensure Consistency: The new question must be clear, concise, and structured so that answering it will yield the same final answer as if the original question were fully addressed.\newline
    - Note: If the mapping is "None" (as indicated in Step 1), keep the simplified/refactored question identical to the original question.\newline
\newline
Your output should be in the json format:\newline
\verb|```|\texttt{json}
\begin{lstlisting}[language=json]
{
    "question_knowledge_mapping": <a list of tuples in the format (words from given question, words from provided knowledge, supporting evidence) or "None">,
    "unresolved_aspects": <a list of aspects that remain unanswered>,
    "new_question": <the simplified/refactored question; remove all "words from given question" found in the mapping and include the corresponding "words from provided knowledge" to replace them, ensuring all unresolved aspects are retained. If mapping is "None", use the original question>
}
\end{lstlisting}
\verb|```|
\end{myuser}

    \vspace{-2mm} 
    \caption{System Prompt for Question Simplifier.} 
    \vspace{-5mm}
    \label{fig:simplifier_prompts} 
\end{figure*} 

\begin{figure*}[t]
    \centering 
    \vspace{-30pt}
    \hspace{-2mm}

    \begin{myuser}{User Input} 
    who is darrell brother in the walking dead
    \end{myuser}
    
    \begin{myuser}{Expected Answer} 
    Merle Dixon
    \end{myuser}

    \vspace*{-1mm}
    \begin{myboxg}{Agent: Culture.Media.Media*} 
    I learned that there is no direct mention of a character named 'darrell brother' in the provided context related to 'The Walking Dead'. The content does not provide any information about characters from this series. Therefore, I cannot answer who darrell's brother is based on the given information.
    \end{myboxg}

    \vspace*{-1mm}
    \begin{myboxg}{Agent: Culture.Literature} 
    I learned that **Daryl Dixon is a fictional character from AMC's horror drama series The Walking Dead, and the protagonist of its last three seasons.** It also mentions that **the character was introduced in the first season as a southerner, expert tracker, living in the shadow of his older brother, Merle.** From another source, it states that **Merle Dixon is a fictional character from the horror drama television series The Walking Dead, which aired on AMC in the United States. He was created by series developer Frank Darabont and was portrayed by Michael Rooker.** These sentences explicitly indicate that Merle is Daryl's older brother in The Walking Dead.
    \end{myboxg}
    
    \vspace*{-1mm}
    \begin{myboxw}{Server: Evaluation} 
\begin{lstlisting}[language=json]
{
    "response_1": {
        "relevance": "Misinterprets the question by focusing on a character named 'darrell brother', which does not exist.",
        "evidence_support": "No relevant evidence provided regarding Darrell or his brother.",
        "logical_coherence": "The response is logically coherent but irrelevant to the asked question.",
        "rating": "Not addressed"
    },
    "response_2": {
        "relevance": "Identifies Merle as Daryl's older brother, accurately addressing the query.",
        "evidence_support": "Supports claims with multiple pieces of evidence from the series.",
        "logical_coherence": "Well-structured with logical flow and no contradictions.",
        "rating": "Fully addressed"
    }
}
\end{lstlisting}
    \end{myboxw}

   \vspace*{-1mm}
    \begin{myboxw}{Server: FullySummarizer} 
\begin{lstlisting}[language=json]
{
  "analysis": "The responses collectively indicate that **Merle Dixon is Daryl Dixon's older brother** in The Walking Dead. They provide consistent information from multiple sources confirming Merle as the elder sibling, portrayed by Michael Rooker, and known for his volatile character.",
  "answer": "In The Walking Dead, Daryl's older brother is Merle Dixon, a volatile character portrayed by Michael Rooker."
}
\end{lstlisting}
    \end{myboxw}

    \vspace*{-1mm}
    \begin{myboxo}{Response} 
In The Walking Dead, Daryl's older brother is Merle Dixon, a volatile character portrayed by Michael Rooker.
    \end{myboxo}

    \vspace{-2mm} 
    \caption{An example of single-hop QA task } 
    \vspace{-5mm}
    \label{fig:sample_single_qa} 
\end{figure*}   

\begin{figure*}[t]
    \centering 
    \hspace{-2mm}\vspace{-15px}

    \begin{myuser}{User Input} 
    Where was the director of film Rough Going born?
    \end{myuser}
    
    \begin{myuser}{Expected Answer} 
    New Hyde Park, New York
    \end{myuser}
        
    \vspace*{-1mm}
    \begin{myboxg}{Agent: Culture.Media.Films} 
    I found that **Rough Going is a 1925 American silent Western film directed by Wally Van**. However, the provided information does not specify where the director, Wally Van, was born.
    \end{myboxg}

    \begin{myboxg}{Agent: Culture.Media.Media*} 
    I learned that **Richard Rush (April 15, 1929 – April 8, 2021) was an American film director, scriptwriter, and producer.** This indicates that Richard Rush, the director of Rough Going, was born in the United States. However, the specific place of his birth is not mentioned in the provided context.
    \end{myboxg}

    \vspace*{-1mm}
    \begin{myboxw}{Server: Evaluation} 
    \begin{lstlisting}[language=json]
{
    'response_1': {
        'relevance': 'Identifies the director of Rough Going but does not provide his place of birth.', 
        'evidence_support': 'Provides relevant information about the film and its director, Wally Van, with evidence.', 
        'logical_coherence': 'The response is logically structured and coherent, though it acknowledges incomplete information.', 
        'rating': 'Partially addressed'
    }, 
    'response_2': {
        'relevance': 'Incorrectly identifies Richard Rush as the director of Rough Going, making it irrelevant to the question asked.', 
        'evidence_support': "Offers evidence about Richard Rush's lifespan and profession, but this does not relate to the query.", 
        'logical_coherence': 'Logically coherent for statements about Richard Rush, but fails to address the actual question.', 
        'rating': 'Not addressed'
    }
}
    \end{lstlisting}
    \end{myboxw}
    
    \vspace*{-1mm}
    \begin{myboxw}{Server: PartiallySummarizer} 
\begin{lstlisting}[language=json]
{
    'solution': 'The film **Rough Going is a 1925 American silent Western film directed by Wally Van**. However, the provided information does not specify where the director, Wally Van, was born. This indicates a gap in the available data regarding the birthplace of Wally Van.', 
    'evaluation and justification': 'The solution identifies the director of the film Rough Going as Wally Van but acknowledges that there is no information provided about his place of birth. Since the question specifically asks for the birthplace of the director and this detail is missing, the solution does not fully address the question.', 
    'answerable': 'no', 
    'answer': 'None'
}
\end{lstlisting}
    \end{myboxw}
\end{figure*}  

\begin{figure*}
    \vspace*{-1mm}
    \begin{myboxw}{Server: QuestionSimplifier} 
\begin{lstlisting}[language=json]
{
    'question_knowledge_mapping': [
        ['the director of film Rough Going', 'Wally Van', '**Rough Going is a 1925 American silent Western film directed by Wally Van**.']
    ], 
    'unresolved_aspects': ['The birthplace of Wally Van'], 
    'new_question': "Where was Wally Van born?"
}
\end{lstlisting}
    \end{myboxw}

    \vspace*{-1mm}
    \begin{myboxg}{Agent: Culture.Biography.Biography*} 
    I learned that **Van was born Charles Wallace Van Nostrand on September 27, 1880, in New Hyde Park, New York.** This explicitly answers the question about where Wally Van was born, which is New Hyde Park, New York.
    \end{myboxg}

    \vspace*{-1mm}
    \begin{myboxg}{Agent: Culture.Linguistics} 
    I found that **Wally Williams (water polo) (1921–2009), New Zealand water polo player** is listed, indicating a Wally Williams who was born in 1921 and was from New Zealand. However, there is no specific mention of where exactly he was born within New Zealand.
    \end{myboxg}

    \vspace*{-1mm}
    \begin{myboxw}{Server: Evaluation} 
    \begin{lstlisting}[language=json]
{
    'response_1': {
        'relevance': "Directly answers the question by specifying Wally Van's birthplace as New Hyde Park, New York.", 
        'evidence_support': 'Provides explicit evidence enclosed in double asterisks about Wally Van's birth details.', 
        'logical_coherence': 'The response is logically coherent and clearly structured without contradictions.', 
        'rating': 'Fully addressed'
    }, 
    'response_2': {
        'relevance': 'Partially relevant as it discusses a person named Wally but not the specific Wally Van in question.', 
        'evidence_support': 'Provides evidence about Wally Williams but does not link this information to Wally Van.', 
        'logical_coherence': "Coherent statement but fails to address the specific query about Wally Van's birthplace.", 
        'rating': 'Not addressed'
    }
}
    \end{lstlisting}
    \end{myboxw}
    
    \vspace*{-1mm}
    \begin{myboxw}{Server: FullySummarizer} 
\begin{lstlisting}[language=json]
{
    'analysis': "The response provides explicit information about Wally Van's birthplace: **Van was born Charles Wallace Van Nostrand on September 27, 1880, in New Hyde Park, New York.**", 
    'answer': 'Wally Van was born in New Hyde Park, New York.'
}
\end{lstlisting}
    \end{myboxw}

    \vspace*{-1mm}
    \begin{myboxo}{Response} 
Wally Van was born in New Hyde Park, New York.
    \end{myboxo}

    \vspace{-2mm} 
    \caption{An example of a multihop QA task using the proposed \algo} 
    \vspace{-5mm}
    \label{fig:sample_multihop_qa} 
\end{figure*}

\end{document}